\documentclass[aps,twocolumn,showlabels,showrefs,amsmath,amssymb,pre,superscriptaddress,floatfix,colors]{revtex4-2}
\usepackage{graphicx,epsfig}
\usepackage{subfigure}  
\usepackage{amsmath}
\usepackage{amsfonts}
\usepackage{xfrac}
\usepackage{enumitem}
\usepackage{xcolor}
\usepackage{color}
\usepackage{epstopdf}
\usepackage[hidelinks]{hyperref}
\usepackage[capitalise]{cleveref}
\usepackage{bbold}
\usepackage{fancyhdr}

\newcommand{\Da}{\mathcal{D}}
\newcommand{\Ha}{\mathcal{H}}
\DeclareMathOperator{\Tr}{Tr}

\begin{document}

\title{Phase Separation of Self-propelled Disks with Ferromagnetic and Nematic  alignment}


\date{\today}

\author{Elena Ses\'{e}-Sansa}
\affiliation{CECAM,  Centre  Europ\'een  de  Calcul  Atomique  et  Mol\'eculaire, Ecole  Polytechnique  F\'ed\'erale  de  Lausanne (EPFL),  Batochime,  Avenue  Forel  2,  1015  Lausanne,  Switzerland}\email{elena.sesesansa@epfl.ch} 
\author{Demian~Levis} 
\affiliation{Departament de Fisica de la Materia Condensada, Universitat de Barcelona, Marti i Franques 1, 08028 Barcelona, Spain}
\affiliation{UBICS  University  of  Barcelona  Institute  of  Complex  Systems,  Mart\'{\i}  i  Franqu\`es  1,  E08028  Barcelona,  Spain}

\author{Ignacio~Pagonabarraga} 
\affiliation{CECAM,  Centre  Europ\'een  de  Calcul  Atomique  et  Mol\'eculaire, Ecole  Polytechnique  F\'ed\'erale  de  Lausanne (EPFL),  Batochime,  Avenue  Forel  2,  1015  Lausanne,  Switzerland}
\affiliation{Departament de Fisica de la Materia Condensada, Universitat de Barcelona, Marti i Franques 1, 08028 Barcelona, Spain}
\affiliation{UBICS  University  of  Barcelona  Institute  of  Complex  Systems,  Mart\'{\i}  i  Franqu\`es  1,  E08028  Barcelona,  Spain}

\begin{abstract}
We present a comprehensive study of a model system of repulsive self-propelled disks in two dimensions with  ferromagnetic and nematic velocity alignment interactions. 
We characterize the phase behavior of the system as a function of the alignment and self-propulsion strength, featuring orientational order for strong alignment and Motility-Induced Phase Separation (MIPS) at moderate alignment but high enough self-propulsion.  
We derive a microscopic theory for these systems yielding a close set of hydrodynamic equations from which we perform a linear stability analysis of the homogenous disordered state. This analysis predicts MIPS in the presence of aligning torques. The nature of the continuum theory allows for an explicit quantitative comparison with  particle-based simulations, which consistently shows that ferromagnetic alignment fosters phase separation, while nematic alignment does not alter either the nature or the location of the instability responsible for it. In the ferromagnetic case, such behavior is due to an increase of the imbalance of the number of particle collisions along different orientations, giving rise to the self-trapping of particles along their self-propulsion direction. On the contrary, the anisotropy of the pair correlation function, which encodes this self-trapping effect, is not significantly affected by nematic torques. Our work shows the predictive power of such microscopic theories to describe  complex active matter systems with different interaction symmetries and sheds light on the impact of velocity-alignment interactions in Motility-Induced Phase Separation. 
\end{abstract}

\maketitle


\section{Introduction}
Inspired by living organisms, active matter made of self-propelled units comprises a wide variety of systems whose common feature is the continuous consumption of energy converted into directed motion. Several well known examples of active systems can be found in the biological world as well as synthetically realized in man-made systems  across scales \cite{Bechinger2016}: examples range from flocks of birds \cite{Bialek2012},  bacteria \cite{Sokolov2007,Zhang2010, Nishiguchi2017,Beer2019}, cells \cite{Kessler1986, Saw2017, Cerbino2017} or cytoskeletal components \cite{Schaller2010,Sumino2012, Inoue2015, DogicRev} to self-propelled colloids \cite{Buttinoni2013, Bricard2013, Ginot2015} or grains \cite{Deseigne2010, Narayan2007, Scholz2018}. 
One of the main interests of these systems lies in the fact that, since they are intrinsically out-of-equilibrium and host different kinds of complex interactions, they display a wide range of emergent collective phenomena. 
 For instance, particle's aggregation in the absence of attractive interactions \cite{Theurkauff2012,Buttinoni2013,Ginot2015,Ginot2018,Geyer2019} and
 the emergence of collective motion \cite{Deseigne2010,Schaller2010,Zhang2010,Sokolov2007,Geyer2019}, are among the most salient examples.  
 
The description of these non-equilibrium phenomena has attracted a great deal of theoretical work over the last decades \cite{MarchettiRev, WinklerRev}. Much progress has been achieved through the study of minimal models capturing some key, hopefully generic, features of active systems. Among them, so-called 'dry models' have played (and are still playing) an important role in the development of a theoretical framework to understand and classify different collective behavior observed in active systems \cite{ChateRev, HaganRev}. Dry models, as their name suggests, neglect the role played by the surrounding medium hosting these self-propelled components, besides as a source of fluctuations and dissipation. They are based on minimal symmetry and dimensionality considerations, following a trend of ideas surely inspired by the theory of critical phenomena. 
Following the symmetry of their constituents, dry models can be classified in different universality classes. For active matter, however, an extra ingredient with no equivalent in the theory of equilibrium critical phenomena has to be taken into account: the symmetry associated to the self-propulsion mechanism itself. In this work, we focus on particles carrying an orientation which sets their self-propulsion direction. Thus, since one can associate an 'arrow' to each particle, they are said to be \emph{polar}. On top of that, one has to consider, as usual, the symmetry of their interactions.

\begin{figure}
	\centering
\includegraphics[width=0.9\columnwidth]{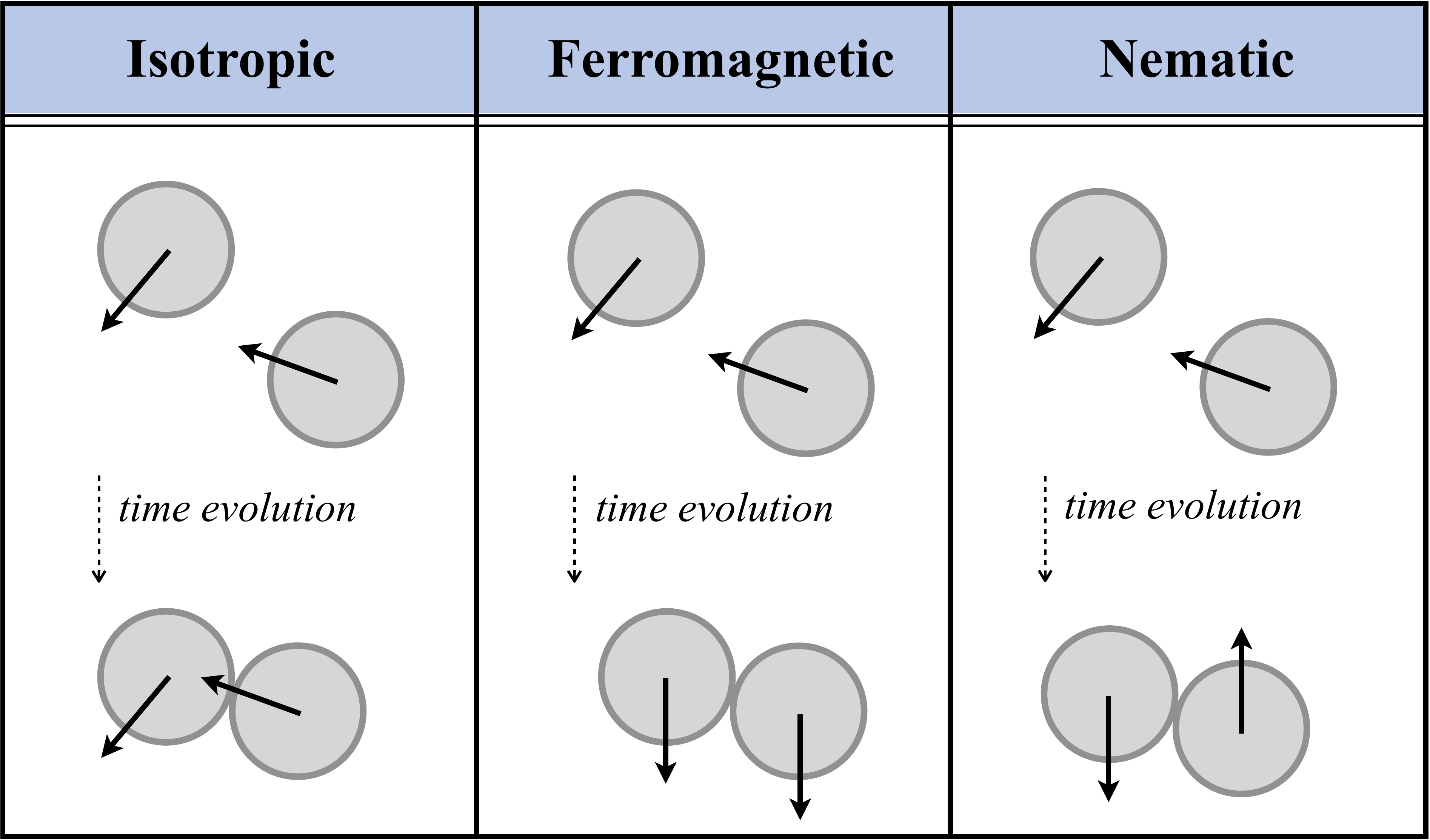}
	\caption{Schematic representation of our model system, comprising Active Brownian Particles subjected to (i) only isotropic excluded volume interactions (first column), (ii) ferromagnetic velocity-alignment  (second column) and (iii) nematic velocity-alignment (third column).}
	\label{particle-based_fig:1}
	\end{figure}

 A natural, and extensively studied, class of active systems made of polar self-propelled particles is the one defined by \emph{isotropic} particle-particle interactions. This encompasses the Active Brownian Particle (ABP) model \cite{Erdmann2000,TenHagen2011,Romanczuk2012}, which describes agents performing a persistent walk and interacting solely through volume-exclusion (as illustrated in the first column \cref{particle-based_fig:1}). Together with its coarse-grained theories \cite{Bialke2013, Speck2014, Speck2015, Wittkowski2014, Nardini2017, Matteo2020, Wittowski2020, CatesField}, this class of models describes the aggregation of self-propelled particles in the absence of attractions, resulting in a macroscopic phase separation at high enough density and self-propulsion strength: the so-called Motility-Induced Phase Separation (MIPS) \cite{Cates2015, TailleurCates2008, Cates_runtumble2013, Fily2012, Redner2013, Stenhammar2014, Winkler2014, Levis2017, Solon2018, Digregorio2018, ClaudioPRL, Lee2019}.
 

 The arguably most studied and first introduced 'universality class' in active matter is the one 
  that considers \emph{ferromagnetic} (or polar) interactions, meaning, particles that locally align their velocity with the one of their neighbors. The celebrated Vicsek model \cite{Vicsek1995} and its continuous descriptions pioneered by Toner and Tu \cite{toner1995}, describing the emergence of collective motion, belong to this class (second column \cref{particle-based_fig:1}).   
Many extensions of the Vicsek model have been considered, accounting for different kinds of alignment rules \cite{ChateRev}. In particular, the case in which particles align along a preferred axis with head-tail \emph{nematic} symmetry has received considerable attention \cite{ginelli2010, Peshkov2012, Bertin2015, Chate2021}, as being models of self-propelled elongated objects, such as most swimming bacteria (third column \cref{particle-based_fig:1}). This class of active systems is typically referred to as \emph{active rods} \cite{PeruaniRev}, both for self-propelled particles with an alignment rule \`a la Vicsek, and for rigid elongated objects for which nematic alignment results from collisions, i.e. anisotropic excluded volume interactions \cite{Peruani2006, abkenar2013, Peruani2015, Shi2018, Dijkstra2019,Jayaram2020, Grossmann2020}. Despite sharing the same nomenclature, self-propelled rods interacting via volume exclusion feature different emergent states than the simplified polar point-like particles with nematic alignment. They can, for instance, form coherently moving polar clusters while Vicsek-like particles cannot, and their aspect ratio turns out to be a crucial parameter as it allows them to go from an alignment dominated regime, to an isotropic regime exhibiting MIPS. 

Here, we investigate the interplay between excluded volume and velocity alignment. 
Most recent studies addressing this problem focus on the role played by the particles' shape in the emergence of different states, mainly looking at how MIPS destabilizes in favor of oriented structures \cite{Shi2018, Dijkstra2019, Jayaram2020, Grossmann2020}. 
 In order to  disentangle the role played by each one of these two interaction mechanisms, short range repulsion and alignment, we consider a system of self-propelled particles with isotropic excluded volume interactions but anisotropic aligning torques, both of ferromagnetic and nematic nature, as illustrated in \cref{particle-based_fig:1}. In other words, we aim at understanding (i) how (isotropic) excluded volume interactions affect the collective behavior of the ferromagnetic and nematic Vicsek-like class of systems, without interfering with more complex aspects related to the shape of the particles and (ii) how ferromagnetic and nematic alignment affect the MIPS of self-propelled disks.  
Such questions have recently been addressed for ferromagnetic  \cite{Peruani2011, Farrell2012,Barre2015,Martin-Gomez2018,Sese-Sansa2018,Geyer2019,VanderLinden2019,Stark2021}, nematic \cite{Bhattacherjee2019} and other kinds of more complex aligning mechanisms present in colloidal experiments \cite{Zhang2020}. However, a unified framework allowing to unravel the impact of aligning torques with different symmetries on the phase behavior of ABP is still lacking. 
 
The paper is organized as follows. In \cref{particlebased_model}, we introduce a microscopic ABP model combining excluded volume and velocity alignment interactions.  In \cref{phase_behavior}, we discuss the phase diagram of the system  subjected to either ferromagnetic or nematic alignment obtained from Brownian dynamics simulations. In \cref{continuum_model}, we systematically derive a continuum description by explicitly coarse-graining the stochastic particle dynamics. 
We then perform a linear stability analysis of the resulting effective hydrodynamic equations, focusing on the stability of the homogeneous isotropic state. 
Finally, we compare the numerical results to the predictions of the analysis of the continuum equations and discuss how a ferromagnetic and a nematic coupling affect the phase separation of repulsive self-propelled disks.

\section{Microscopic ABP model} \label{particlebased_model}

We consider a 2D system of $N$ particles in a $L\times L$ box with periodic boundary conditions  at positions $\textbf{r}_{i}$ and with orientations $\textbf{e}_{i} = \left(\cos \varphi_i,\sin \varphi_i \right)$, whose dynamics are governed by the set of coupled over-damped Langevin equations:
\begin{equation}\label{modelmeth_eq:1}
\dot{\textbf{r}}_{i}(t) = v_{0}\textbf{e}_{i} + \mu \textbf{F}_{i} \left( \left\{ \textbf{r}_j (t)\right\} \right)+ \sqrt{2D_{0}}\boldsymbol{\eta}_{i}(t)
\end{equation}
\begin{equation}\label{modelmeth_eq:2}
\dot{\varphi}_{i}(t) = - \mu_r\frac{\partial \Ha}{\partial \varphi_i}+ \sqrt{2D_{r}}\nu_{i}(t) 
\end{equation}
Particles self-propel at constant speed $v_0$ in the direction given by $\textbf{e}_{i}$. The time evolution of $\textbf{e}_{i}$ is subjected to local torques $T_i= - \frac{\partial \Ha}{\partial \varphi_i}$ which derive from a Vicsek-like alignment rule that we specify below. Particles also interact through isotropic, short-range, repulsive forces, $\textbf{F}_{i}$.
The system is in contact with a thermal bath, modeled by $\boldsymbol{\eta}_{i}$, a Gaussian white noise with zero mean and unit variance. Orientations are subjected to rotational noise, $\nu_{i} $, also described by a Gaussian white noise with zero mean and unit variance. Rotational noise sets a characteristic  time scale, given by the inverse of the rotational diffusion coefficient, $\tau_{r} = D_r ^{-1}$, the persistence time, and a characteristic length scale given by $l_p=v_0\tau_r$, the persistence length. The thermal diffusion coefficient $D_{0}$ and the mobility $\mu$ fulfill the Einstein relation $D_{0}=\mu k_{B} T$, where $k_{B}$ is the Boltzmann constant and $T$ the temperature of the bath. 

We now specify the different interaction terms, included in our model equations (\cref{modelmeth_eq:1,modelmeth_eq:2}) and used in the present work.

\subparagraph*{Excluded volume} 
The excluded volume interaction, $\textbf{F}_{i} = - \sum_{j \neq i} \nabla_i u \left(r_{ij} \right)$, derives from a WCA potential,
\begin{equation}
u (r_{ij}) =\begin{cases} 4 u_0 \left[ (\frac{\sigma}{r_{ij}})^{12} - (\frac{\sigma}{r_{ij}})^6 \right] + u_0 & r_{ij} \leq R \\
0 & r_{ij} > R
\end{cases}
\label{modelmeth_eq:3}
\end{equation}
where $r_{ij}=|\bold{r}_i-\bold{r}_j|$.
The cutoff distance is $R=2^{1/6} \sigma$ and $u_0$ corresponds to the characteristic energy scale of the potential.

\subparagraph*{Aligning torques } 
Local torques derive from the following Hamiltonian (potential)
\begin{equation}
\Ha =- \sum_{i,j}\, v(r_{ij}) w(\varphi_{ij})
\label{modelmeth_eq:4}
\end{equation}
where $\varphi_{ij}=\varphi_i-\varphi_j$ and 
 \begin{equation}
v(r_{ij}) = \begin{cases} \frac{2}{\pi R^2_{\varphi}} (r_{ij}-R_{\varphi})^2 &  r_{ij} < R_{\varphi}  \\ 
0 &r_{ij} > R_{\varphi}
\end{cases}
\label{modelmeth_eq:5}
\end{equation}
is a spatially decaying function with a cutoff distance $R_{\varphi}$,  setting the interaction range beyond which particles do not align.
The spatial dependency of $v\left(r_{ij} \right)$ ensures that there are no discontinuities in the resulting torque. The angular dependency of the alignment potential, $w\left(\varphi_{ij} \right)$, is chosen in different ways in order to study  interactions of different nature: (a) ferromagnetic and (b) nematic alignment.

In the absence of alignment interactions, the model described by \cref{modelmeth_eq:1,modelmeth_eq:2} reduces to the Active Brownian Particle model. In this limit, the system undergoes Motility-Induced Phase Separation upon increasing the self-propulsion speed. Furthermore, in the absence of excluded volume interactions, $u_0 =0$, the model describes point-like particles \emph{\`a la} Vicsek, subjected to alignment interactions of different nature. 
\\

\paragraph{Ferromagnetic alignment }
Ferromagnetic torques lead to a Langevin variant (in continuous time) of the Vicsek model, which, in the limit of vanishing velocities reduces to the equilibrium 2D XY model of a ferromagnet. Indeed, a ferromagnetic coupling in \cref{modelmeth_eq:2} can be derived from the 2D XY Hamiltonian
 \begin{equation}
\Ha = -J \sum_i \sum_{j \in \omega_i} v(r_{ij})\, \textbf{e}_i \cdot  \textbf{e}_j
\label{modelmeth_eq:6}
\end{equation}
meaning,
\begin{align}
&w\left(\varphi_{ij} \right) = J \cos \left(\varphi_{ij} \right)
\end{align} 
where $\omega_i$ is the vicinity of particle $i$, defined by $R_{\varphi} $ and $J>0$ is the coupling strength. 
\\

\paragraph{Nematic alignment }
 Nematic torques tend to align the direction of self-propulsion of neighboring particles along the same axis but with no head-tail preference, see \cref{particle-based_fig:1}, in the same fashion as uniaxial nematic liquid crystals \cite{Andrienko2018}. Thus, the nematic interaction can be modeled by the following Hamiltonian
\begin{equation}
\Ha =-  J\sum_i \sum_{j \in \omega_i} v(r_{ij})\,  \textbf{q}_i \cdot \textbf{q}_j
\label{modelmeth_eq:6}
\end{equation}
where $\textbf{q}_i= \textbf{e}_i \otimes \textbf{e}_i - \frac{1}{2} \mathbb{1}$ is the nematic tensor \cite{deGennes}. Note that  \cref{modelmeth_eq:6}  has the same structure as the 2D XY model interaction, but now the nematic tensor plays the role of the orientation of the particles (spins) \cite{Jef2011}. 
From such Hamiltonian one gets  (see \cref{deriv_nemat_hamilt})
\begin{align}
&w\left(\varphi_{ij}\right) = J\cos\left(2\varphi_{ij} \right)
\end{align}

At this stage, it is possible to identify the relevant set of dimensionless parameters:
the average packing fraction 
\begin{equation}
\phi=\frac{N \pi R^2}{4 L^2}\equiv \overline{\rho}\frac{\pi R^2}{4}\,,
\end{equation} 
the reduced coupling parameter 
\begin{equation}
g = \frac{2\mu_rJ}{\pi R^2_{\varphi}D_r}\,,
\end{equation} 
 accounting for the strength of the alignment interaction as compared to the rotational diffusion, and the P\'eclet number 
 \begin{equation} 
 \mathrm{Pe} = \frac{v_0}{R D_r}=l_p/R
 \end{equation} 
 quantifying the persistence of the particle's motion. 
\\

We systematically study the system's phase behavior fixing $\phi=0.4$ and varying $g$ and $\mathrm{Pe}$. To this end, we perform Brownian dynamics simulations of the model described in \cref{modelmeth_eq:1,modelmeth_eq:2} with $N=4000-16000$ particles in a  $L \times L$ box subjected to periodic boundary conditions (PBC). 
The steric interaction cutoff distance is set to $R=1$, which in turn gives the unit of length (interpreted as the effective diameter of the particles). The strength of the pairwise repulsive interaction is $u_0 = 100$, in units of the thermal energy, $k_BT=1$. The time unit is given by $\tau_0 = R^2/D_0 = 1$, where the thermal diffusion coefficient is $D_0=R^2D_r/3$, with fixed rotational diffusion coefficient $D_r = 3/\tau_{0}$. This, in turn, sets the characteristic decorrelation time, $\tau_r = D^{-1}_r$. The alignment cutoff distance is set to $R_{\varphi} = 2 R$, so that particles need not be in contact in order to mutually align their directions of self-propulsion. We explore a range of $v_0$ values that goes from $0$ to $400$, corresponding to $\mathrm{Pe} \in \left[0, 133.3\right]$. The coupling constant $J$ takes values from $0$ to $320$ and thus $g \in \left[0, 17.0 \right]$. Finally, mobilities are set to $\mu =\mu_r= 1$.

The simulations are performed by initializing the system in a random configuration and letting it relax to its steady-state. The results reported are obtained by averaging over ensembles of 1000 independent configurations at the stationary state (in the regime of low $\mathrm{Pe}$ where density and polarization fluctuations are bigger, we sample systems of $N=8000$ and average over 5000 independent configurations). We use an Euler-Mayurama algorithm to integrate the equations of motion. The time-step employed to discretize the equations of motion ranges from $\Delta t = 10^{-5}$ to $\Delta t = 2 \cdot 10^{-6}$, depending on the value of Pe. Each simulation lasts for $320\tau_r$, in terms of the decorrelation time $\tau_r$. Initially, we let the system evolve for $120\tau_r$, until it reaches the steady-state, and then record a configuration every $2\tau_r$ for further statistical analysis.

	\begin{figure*}
	\centering
	\includegraphics[trim=75 120 0 0,width=\textwidth]{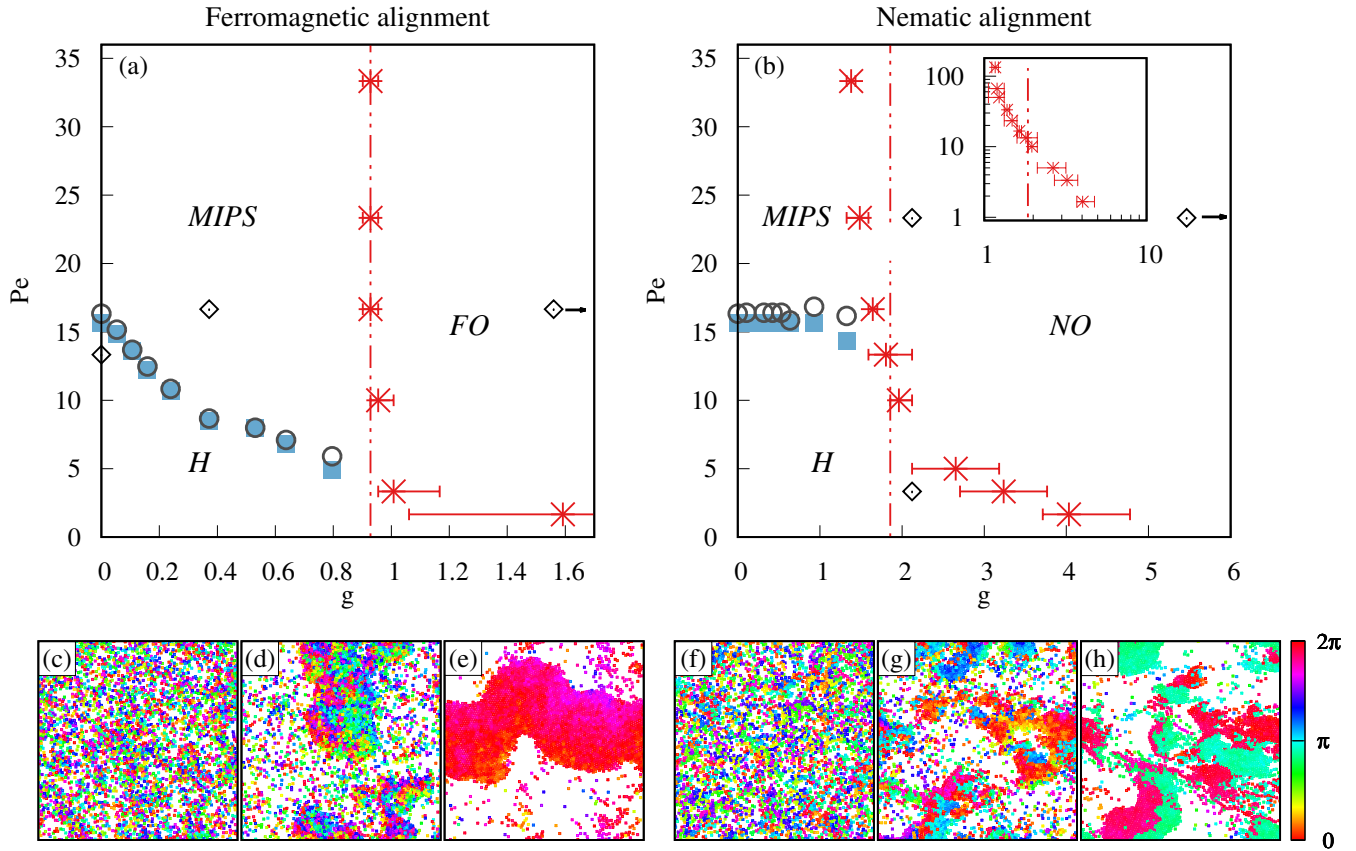}
	\caption{Phase diagram in the $(g, \mathrm{Pe})$ plane for a system at $\phi=0.4$ with (a) local ferromagnetic alignment and (b) local nematic alignment. The phase separation from a homogeneous (\textit{H}) to a MIPS state is indicated by blue squares. The prediction of the phase separation given by the continuum theory is marked by grey circles. The onset of ferromagnetic order (FO) and nematic order (NO) is marked by red symbols with error bars. The inset in (b) shows the onset of nematic order up to a value of $\mathrm{Pe}=133.4$ in log-log scale. 
Broken lines correspond to $g=g_f=0.93$ in (a) and $g=2g_f$ in (b).  	
	Snapshots of the system at different points of the phase diagrams (indicated by symbols) (c): (0, 13.3); (d): (0.37, 16.6); (e): (7.96, 16.6) in the ferromagnetic case; (f): (2.12, 3.3); (g): (2.12, 23.4); (h): (13.23, 23.4) in the nematic case. Particles are colored according to their self-propulsion direction, using a cyclic color code. Particles with the same color share the same orientation. }
	\label{phasediag_fig:1}
	\end{figure*}
We explored by means of numerical simulations the behavior of the system by varying $g$ and Pe, subjected to either ferromagnetic or nematic alignment at fixed $\phi = 0.4$. The resulting  state diagram obtained for systems with $N=4000$ is shown in \cref{phasediag_fig:1}.
Aligning torques induce the emergence of orientational order in the system, either of ferromagnetic or nematic nature, while the competition between self-propulsion and excluded volume interactions triggers phase separation. The study of the latter in the presence of aligning interactions is the main object of the present article and presented in sections \ref{continuum_model} and \ref{comparison}.

\section{Phase behavior} \label{phase_behavior}

\subsection{Emergence of orientational order}

Upon increasing the tendency of particles to align at fixed self-propulsion velocity, the system undergoes a phase transition between an isotropic and an oriented state (see snapshots \cref{phasediag_fig:1}).
At low values of the coupling parameter, the orientational dynamics of particles is dominated by rotational noise, thus leading to a  disordered state. Upon increasing $g$, local alignment torques overcome rotational noise and eventually trigger the emergence of global orientational order. A macroscopic fraction of the particles in this state is thus aligned, leading to the emergence of collective motion (or flocking), \cref{phasediag_fig:1} (e) and (h).


Obviously, the nature of the ordered phases strongly depends on the symmetry of the alignment interactions. For ferromagnetic alignment, the ordered state displays polar order and strong density heterogeneities in the form of lanes, as illustrated in \cref{phasediag_fig:1} (e), along which a macroscopic fraction of the system moves (roughly) along the same direction. The formation of dense structures such as traveling bands are typical in flocking Vicsek-style models \cite{ChateRev}. Extensions of such models  including volume interactions display richer structures, and among them, it is typical to find lanes \cite{Peruani2011,Farrell2012, Martin-Gomez2018}. 

Local nematic torques lead, instead, to an ordered state where particles self-propel along the same axis but in opposite directions, \cref{phasediag_fig:1} (h). Here,  the ordered phase is also characterized by the formation of dense structures, although of different nature than in the ferromagnetic case.  Particles with nematic alignment aggregate  into coherently-moving structures, with a high degree of local polar order. These domains collide and interpenetrate, giving rise to elongated, nematically ordered structures that we call, by extension of Vicsek-type models, nematic bands. An illustration of these nematic bands is provided by the snapshots \cref{phasediag_fig:1} (g)-(h). 
While nematic bands are generically found in Vicsek-type models of active rods \cite{ChateRev}, the situation becomes quite more intricate when excluded volume interactions are considered. In active rod models, made of elongated self-propelled particles that align through collisions, one generically observes the formation of polar lanes, similar to the ones we observe in the ferromagnetic case, at high enough densities and shape anisotropy \cite{abkenar2013,  Shi2018,  Jayaram2020, Grossmann2020}. Here, we also find a tendency towards polar ordering, however, only locally. The isotropic shape of our particles does not allow for the establishment of a large-scale polar structure even for very strong coupling. 

\begin{figure}
	\centering
	\includegraphics[trim=110 70 0 230,width=\columnwidth]{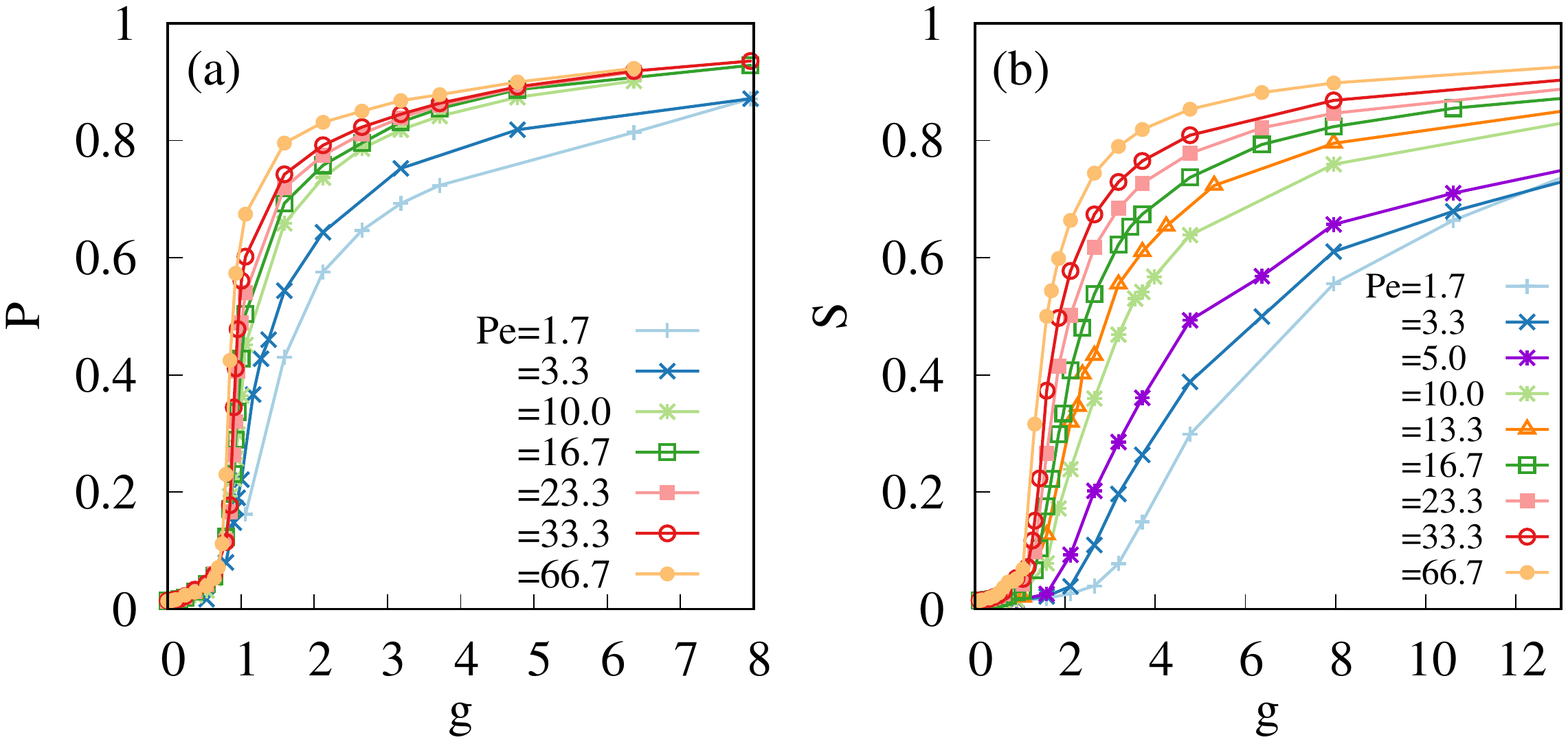}
	\caption{(a) Global polarization as a function of the normalized ferromagnetic alignment strength for different values of the self-propulsion speed; (b) Global nematic order as a function of the normalized nematic alignment strength.}
	\label{phasediag_fig:2}
	\end{figure}

The emergence of ferromagnetic order is characterized by a non-zero value of the polarization 
 \begin{equation}
 P = \left| \frac{1}{N}  \sum_{i=1}^N \textbf{e}_{i} \right|\,,
 \label{onset_orientorder-fig:1}
 \end{equation}
 while nematic order is quantified by the scalar nematic order parameter $S$ \cite{Chate2006}, defined as, 	
	\begin{equation}
	\begin{split}
	S =  \frac{2}{N}\left[\left(\sum_{i} \cos^2 \varphi_i  -\frac{1}{2}\right)^2 + \left( \sum_{i} \cos \varphi_i \sin \varphi_i \right)^2 \right]^{\frac{1}{2}}\,.
	\label{onset_orientorder-fig:2}
	\end{split}
	\end{equation}
The evolution of the averaged order parameters as a function of the coupling strength is shown in \cref{phasediag_fig:2} (a) and (b) in the ferromagnetic and nematic case, respectively. 
To locate the onset of global ferromagnetic order shown in red symbols in the phase diagram, \cref{phasediag_fig:1} (a), we compute the susceptibility $\chi(P) = N (\left<P^{2} \right> - \left<P\right>^2)$, where $\langle \cdot \rangle$ denotes an ensemble average. In the case of nematic alignment, we observe strong fluctuations in $S$ at low $\mathrm{Pe}$, resulting in a broad susceptibility with an ill-defined peak. However, at high values of $\mathrm{Pe}$, $\chi(S)$ peaks at a value of $g$ corresponding to $S=0.2$. It is for this reason that we use the latter threshold value as a criterion to locate the onset of nematic order at any value of $\mathrm{Pe}$. The phase diagram \cref{phasediag_fig:1} (b) shows in red symbols the critical value of $g$ corresponding to $S=0.2$.

From this analysis, we identify the onset of ferromagnetic order at $g_{f} = 0.93$. This critical coupling is largely independent of the value of $\mathrm{Pe}$, see \cref{phasediag_fig:1} (a). The independence of the onset of flocking on Pe has been already reported in numerical simulations \cite{Martin-Gomez2018,Sese-Sansa2018} and mean-field calculations  \cite{Farrell2012}, according to which it is located at $g\approx 0.63$. 

The onset of nematic order $g_n(\mathrm{Pe})$ has a stronger dependence on the value of $\mathrm{Pe}$ for the finite system sizes explored. The higher the $\mathrm{Pe}$ is, the lower the alignment strength has to be in order to trigger nematic order. A Pe-dependence of the onset of nematic order has been recently observed in other simulations of nematically aligning ABP \cite{Bhattacherjee2019}.
The parallel analysis of ferromagnetic and nematic interactions we perform here allows us to identify that,  as Pe increases, the critical value $g_n$  decreases, first approaching $2g_f$, the value expected in the absence of self-propulsion, and then going below that value at very large Pe, saturating at a value around $g_n(\text{Pe}\gg 1)\approx 1.15$. Note that in this high-Pe regime, the persistence length $l_p$ becomes larger than the linear system size $L=88.6$, such that large-scale fluctuations are strongly suppressed. Thus, the behavior at high Pe is likely to be strongly affected by finite size effects and controlled by mean-field behavior. For Pe$>89$, the persistence length of the particles exceeds the linear size of the box $L$. Larger system sizes, such that $L/l_p\ll1$, would be needed in order to analyze finite size effects in this high activity regime. We however did not attempt to characterize the nature of the orientated state. Above Pe$=89$, the observed flocking transition is mean-field like and does not depend anymore on the value of Pe. Instead, for smaller values of Pe, or larger system sizes, the nematic order parameter decays at fixed $g$ for increasing $l_p$.

It is worth mentioning at this stage that the nature of the ordered state in systems of Vicsek particles with nematic alignment has arouse some debate over the last decade. While it is now clear that true long-range order can arise for ferromagnetic interactions, numerical and analytical results have not convincingly yield a conclusion in this respect for the nematic case \cite{toner2005hydrodynamics, ginelli2010}, until very recently \cite{Chate2021}. In this latter work, it is shown that  very large systems need to be explored in order to grasp the asymptotic quasi-long-range order nature of the ordered state. Thus, in finite, yet large, systems of linear size smaller than a characteristic length scale (which can be made very large) nematic order appears to be long-range. The location of the threshold reported  in \cref{phasediag_fig:1} has thus to be taken as a description of the behavior of our finite system, bearing in mind the aim of the present study, which is the understanding of the particle aggregation mechanism in the presence of different kind of aligning interactions. The exploration of the phase diagram presented here responds to the need of setting  the parameters for our subsequent study of phase separation, but not as an attempt of establishing the asymptotic behavior of the system in the $N\to\infty$ limit. In the presence of excluded volume interactions though, the nature of the phase transition triggered by nematic interactions in systems of self-propelled particles remains an open challenge. We do not aim at addressing this question here. 


\subsection{Phase Separation}

In the absence of effective torques, $g=0$, the system undergoes a phase separation from a homogeneous (denoted H) to a phase separated state (MIPS) by increasing Pe, \cref{phasediag_fig:1}. 
At finite $g$ below the emergence of orientational order, there is a region of the phase diagram where the system de-mixes  into a dense region, where particles move slowly, and a dilute region where particles move fast. As shown in \cite{Sese-Sansa2018} for ferromagnetic alignment, such phase separation can be attributed to the MIPS mechanism, meaning, a fast reduction of the particles velocity with increasing local density.

In what follows, we study the impact of alignment (both ferromagnetic and nematic) on the global phase separation induced by motility.   We use the fraction of particles in the largest cluster of the system, $\Pi$, as a phenomenological order parameter to identify the onset of phase separation (see \cref{appx_phase-sep-characteriz} for further details), or spinodal line. We consider the system in a phase separated state when $\Pi>0.08$ and report the  critical value of $\mathrm{Pe}$ obtained in this way in the phase diagrams \cref{phasediag_fig:1} (a) and (b). 

As shown in \cref{phasediag_fig:1} (a), for ferromagnetic alignment, MIPS is shifted to lower values of $\mathrm{Pe}$ as the coupling parameter $g$ is increased, as reported in \cite{Peruani2008, Sese-Sansa2018}. Ferromagnetic alignment enhances the aggregation of particles and the eventual phase separation of the system. 
In contrast, for nematic alignment, the critical self-propulsion speed at which MIPS takes place remains fairly unchanged as the coupling parameter $g$ is increased, as shown in \cref{phasediag_fig:1} (b).

\begin{figure}
	\centering
	 \includegraphics[trim=230 20 130 20,width=\columnwidth]{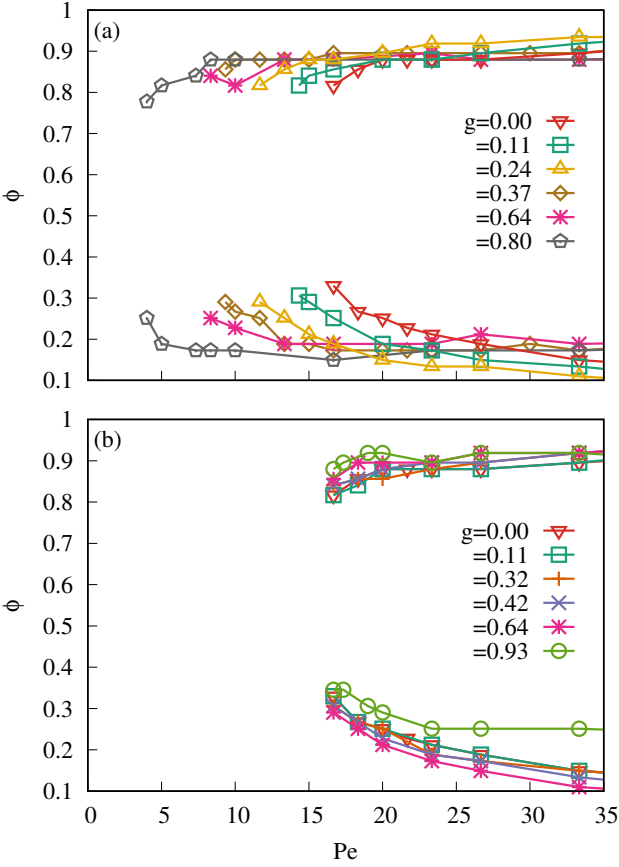}
	\caption{Phase coexistence regions for  several values of  the (a) ferromagnetic and (b) nematic alignment strength, $g$. The coexistence region shifts to lower values of Pe as the ferromagnetic coupling is increased, while increasing the nematic coupling does not significantly affect the coexistence region of non-aligning ABP.}
	\label{binodal}
	\end{figure}

To further characterize the effect of different  alignment interaction on MIPS, we construct the binodals of the system, shown in \cref{binodal}, from the analysis of local density distributions. For ferromagnetic alignment, the region of coexistence is shifted to lower values of Pe as $g$ is increased.  On the contrary, for nematic alignment, the coexistence regions do  not change within our numerical accuracy, in agreement with the results presented in \cref{phasediag_fig:1}. The MIPS coexistence region of ABP is not significantly affected by the presence of nematic alignment interactions. The system's finite size does not have a strong effect in the MIPS characterization. To show the robustness of the results obtained at $N=4000$ we have reproduced the binodals for bigger system sizes ($N=8000,16000$), see \cref{appx_systemsize_fig:1} in \cref{system_size}.

 With the aim of shedding light on the mechanisms triggering MIPS in the presence of aligning interactions, we develop below a continuum theory by explicitly coarse-graining the microscopic ABP dynamics \cref{modelmeth_eq:1,modelmeth_eq:2}. We  then perform a linear stability analysis of the resulting field equations that we can directly compare with our simulation results, thus explaining why ferromagnetic interactions enhance MIPS while nematic ones do not.  

\section{Coarse-grained description} \label{continuum_model}

\paragraph{Derivation of  hydrodynamic equations} The overdamped Langevin dynamics \cref{modelmeth_eq:1,modelmeth_eq:2} can be equivalently described by the following $N$-body Smoluchowski equation \cite{Risken1996, Bialke2013, Speck2014, Speck2015},
\begin{equation}
\begin{split}
\partial_{t} \psi_{N}& = \sum_{i=1}^{N} \nabla_{i} \cdot \left[ (\nabla_{i} U) \psi _{N} - v_{0} \textbf{e}_{i} \psi _{N} + D_0\nabla_{i} \psi _{N} \right]  +\\
 & \qquad  \sum_{i=1}^{N} \partial _{\varphi_{i}} \left[ \left(\partial_{\varphi_{i}} \mathcal{H} \right) \psi _{N}+D_{r} \partial_{\varphi_{i}} \psi_{N} \right]
\label{modeldescrip_eq:0}
\end{split}
\end{equation}
where $\psi_{N}(\Gamma=\{\textbf{r}_{i}, \varphi_{i}\}_{i=1..N},t)$ is the joint probability distribution to find our $N$ particles at a given position with a given  orientation at time $t$. Particles self-propel at constant speed $v_0$ and are subjected to translational and rotational diffusion. From now on we consider $D_0=1$ and  drop it in the following expressions.
Short range interactions are modeled by two independent potentials, describing  excluded volume and alignment interactions,
\begin{equation}
\begin{split}
U(\left\{ \textbf{r}_j \right\})=&\sum _{i=1}^{i=N}  \sum_{i<j} u(|\textbf{r}_{j} - \textbf{r}_{i}|)    \\
\Ha(\left\{ \textbf{r}_j \right\}, \left\{ \varphi_j \right\}) = &\sum_{i<j} v(|\textbf{r}_{j} - \textbf{r}_{i}|)   w(\varphi_{j} - \varphi_{i}) 
 \end{split}
\label{theoret-mean-field_eq:2}
\end{equation}

Assuming the indistinguishability of particles, it is possible to integrate out $(N-1)$ variables, $\psi_{1} (\textbf{r}_{1},\varphi_{1};t) = N \int_{-\infty}^{\infty} d\textbf{r}_{2} ... d\textbf{r}_{N} \int_{0}^{2\pi} d\varphi_{2} ... d\varphi_{N}  \psi_{N}$, to obtain a  Smoluchowski equation for the 1-body distribution,
\begin{equation}
\begin{split}
\partial_{t} \psi_{1} &= - \nabla_{1} \cdot \left[ \textbf{F} \left(\textbf{r}_{1},\varphi_{1};t\right)+ v_{0} \textbf{e}_1 \psi _{1} - \nabla_{1} \psi _{1} \right] \\
&\qquad  - \frac{\partial}{\partial \varphi_{1}} \left[ T \left(\textbf{r}_{1},\varphi_{1};t \right)- D_{r} \frac{\partial \psi_{1}}{\partial \varphi_{1}} \right]
 \end{split}
\label{theoret-mean-field_eq:1}
\end{equation}

The effective force reads, 

\begin{equation}
\begin{split}
\textbf{F} (\textbf{r}_{1},\varphi_{1};t)=\int_{-\infty}^{\infty} d\textbf{r}_{2} \int_{0}^{2\pi} d\varphi_{2} u ' \left(r_{12}\right)\frac{\textbf{r}_{12}}{r_{12}} \psi _{2}
\end{split}
\label{modeldescrip_force_eq:1}
\end{equation}
and the effective torque corresponds to, 
\begin{equation}
\begin{split}
T(\textbf{r}_{1},\varphi_{1};t) = \int_{-\infty}^{\infty} d\textbf{r}_{2}  \int_{0}^{2\pi} d\varphi_{2}  v\left(r_{12}\right) w' \left( \varphi_{12}\right)  \psi _{2}
\end{split}
\label{modeldescrip_force_eq:2}
\end{equation}
where $\textbf{r}_{12} = \textbf{r}_{2} - \textbf{r}_1$ and $\varphi_{12} = \varphi_{2} - \varphi_1$. Note that both $\textbf{F} (\textbf{r}_{1},\varphi_{1};t)$ and $T (\textbf{r}_{1},\varphi_{1};t)$ depend on the two-body probability density, $\psi_{2}(\textbf{r}_{1}, \textbf{r}_{2}, \varphi_{1}, \varphi_{2},t)$, and encode the microscopic interactions exerted by the surrounding particles into the tagged particle (labeled \textit{1}). Eq. (\ref{theoret-mean-field_eq:1}) thus constitutes the first equation of a BBGKY-like hierarchy of coupled equations involving multi-body distribution functions.

\begin{figure}
	\centering
\includegraphics[width=0.7\columnwidth]{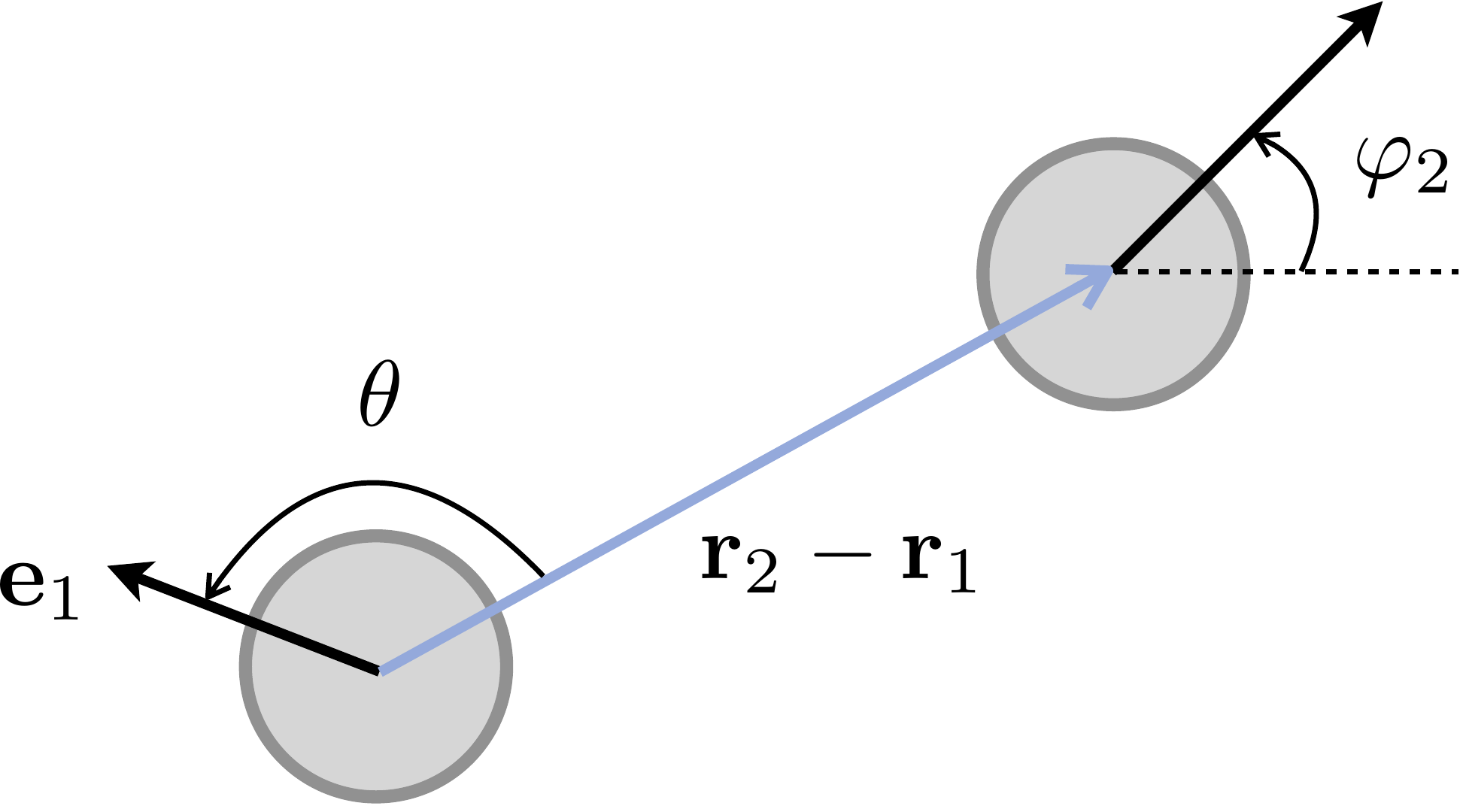}
	\caption{Schematic representation of two Active Brownian Particles located at $\bold{r}_1$ and  $\bold{r}_2$, setting the notations used throughout the paper. $\theta$ is the angle encompassed between the vector distance and the direction of self-propulsion, $\textbf{e}_1 = (\cos \varphi_{1}, \sin \varphi_{1})$.}
	\label{continuum-model_fig:1}
	\end{figure}

We decompose the two-body probability density using the following identity 
\begin{equation}
\psi_{2}(\textbf{r}_1,\textbf{r}_2,\varphi_{1},\varphi_{2},t) =\bar{\rho}\, \psi_{1}(\textbf{r}_1,\varphi_{1},t)\,  \mathcal{G}(r_{12},\theta,\varphi_{12},t)
\label{theoret-mean-field_eq:5}
\end{equation}
where $\bar{\rho}$ is the average density,  $\mathcal{G}(r_{12},\theta,\varphi_{12},t)$ the pair-correlation function and  $\theta$ the angle encompassed between the vector distance $\textbf{r}_{12} $ and the orientation of the tagged particle (see sketch in  \cref{continuum-model_fig:1}). This decomposition allows us to encode all the information on the microscopic structure of the system in the correlation function $\mathcal{G}(r_{12},\theta,\varphi_{12},t)$, interpreted as the probability of finding a particle with orientation $\varphi_{2}$ in the plane-direction $\theta$, at a distance $ r_{12} = |\textbf{r}_{12} |$ from the tagged particle (at $\textbf{r}_{1}$  with orientation $\varphi_{1}$, see Fig. \ref{continuum-model_fig:1}) From now on, we drop the subscripts for clarity.

We then insert the two-body probability density decomposition \cref{theoret-mean-field_eq:5} in \cref{modeldescrip_force_eq:1,modeldescrip_force_eq:2}, and consider the projected effective force in the direction of self-propulsion of the tagged particle, which leads to the expression, 
\begin{equation}
\begin{split}
\textbf{e} \cdot \textbf{F}  =  -\bar{\rho} \psi_{1} \zeta
\end{split}
\label{theoret-mean-field_eq:6}
\end{equation}
where $\zeta$ is a scalar coefficient that reads,
\begin{equation}
\begin{split}
&\zeta=-\int_{0}^{\infty} dr \: ru ' (r ) \int_{0}^{2\pi} d\theta \cos \theta\int_{0}^{2\pi} d\varphi  \mathcal{G}(r,\theta,\varphi,t) 
\end{split}
\label{theoret-mean-field_eq:7}
\end{equation}
Similarly, the torque can be expressed as,
\begin{equation}
\begin{split}
T = -\bar{\rho} \psi_{1} \varepsilon
\end{split}
\label{theoret-mean-field_eq:8}
\end{equation}
where the $\varepsilon$ coefficient reads,
\begin{equation}
\begin{split}
\varepsilon=- \int_{0}^{\infty} dr \:  r v( r) \int_{0}^{2\pi} d\theta  \int_{0}^{2\pi} d\varphi  w'( \varphi) \mathcal{G}(r,\theta,\varphi,t) 
\end{split}
\label{theoret-mean-field_eq:9}
\end{equation}
We have thus recast the dependence on pair-wise correlations in the two coefficients $\zeta$ and $\varepsilon$ (see  \cref{appx_continuum_model_forcetorque} for more details).

{We now write the force $\textbf{F} (\textbf{r}_{1},\varphi_{1};t)$   in the vector basis defined by the direction of self-propulsion, $\textbf{e}$, and the gradient of the probability density, $ \nabla \psi_{1}$, following a Gram-Schmidt  procedure (see \cref{appx_GSortho} for the full derivation),  leading to,
\begin{equation}
\textbf{F} \approx \big(\textbf{e} \cdot \textbf{F} \big )\textbf{e} + \big(1-\Da \big)\nabla\psi_{1}
\label{theoret-mean-field_eq:10}
\end{equation}
where,
\begin{equation}
\Da=1-\textbf{F} \cdot\big[ \nabla\psi_{1} - (\textbf{e} \cdot \nabla\psi_{1})\textbf{e} \big] \frac{1}{|\nabla\psi_{1}|^{2}}
\label{modeldescrip_eq:14bis}
\end{equation} 
This is an approximation as the basis vectors chosen are time dependent and might eventually, although  unlikely, become collinear at some particular time during the evolution.

One can then rewrite the one-body Smoluchowski equation, \cref{theoret-mean-field_eq:1}, inserting the projected force, \cref{theoret-mean-field_eq:6}, as well as the expression of the torque, \cref{theoret-mean-field_eq:8}, to obtain, 
\begin{equation}
\begin{split}
\partial_{t} \psi_{1} = - \nabla \cdot \big[ v_{\bar{\rho}} \,\textbf{e} \psi_{1}- \Da \nabla\psi_{1} \big] +\frac{\partial}{\partial \varphi} \big[ \bar{\rho} \psi_{1} \varepsilon + D_{r} \frac{\partial\psi_{1}}{\partial \varphi} \big]
\end{split}
\label{modeldescrip_eq:14}
\end{equation}
The first term in the right hand side (RHS) stands for translational advection, defining an effective self-propulsion speed 
\begin{equation}
v_{\bar{\rho}} = v_0 -\bar{\rho} \zeta 
\end{equation}
that decays with the mean density $\bar{\rho}$ at a rate given by $\zeta$, which can thus be interpreted as a \textit{translational friction coefficient}, accounting for the arrest of particles in crowded environments and stemming from excluded volume interactions (see \cref{theoret-mean-field_eq:7}).
The second term in the RHS \cref{modeldescrip_eq:14} corresponds to translational diffusion with an effective diffusivity $\Da$. Following \cite{Bialke2013,Speck2015}, we make the assumption that $\Da$ is uniform, corresponding to the long time diffusion coefficient of the system in the passive limit. 
Such approximation of the effective diffusivity Eq. \ref{modeldescrip_eq:14bis} has been justified and put into test in previous works \cite{joakimPRL2013, Speck2015}. The diffusivity from the long-time mean-squared displacement can be written as $D=\Da+v^2(\rho)/2$, such that in the absence of self-propulsion $\Da$ corresponds to the diffusion coefficient of a passive system. This implies that, from now on, all the dependency on pair-wise correlations will be fully encoded in $\zeta$ and $\varepsilon$.
The effect of the third term in the RHS \cref{modeldescrip_eq:14}  is to advect the orientations. 
Similarly to $\zeta$, $\varepsilon$ can be interpreted as a \textit{rotational friction coefficient}. 
The last term accounts for rotational diffusion.

To proceed, we close the hierarchy of coupled equations by considering the effective friction coefficients as constants \cite{Bialke2013}. This is the central approximation of our approach which allows us to construct an effective hydrodynamic description.
We define the first three moments of the one-body probability distribution, $\psi_{1}$, namely, the density field 
\begin{equation}
\rho (\textbf{r},t) \equiv \int_{0}^{2 \pi} d \varphi \psi_{1} (\textbf{r},\varphi,t)\,,
\end{equation}
 the polarization 
 \begin{equation}
 \textbf{p} (\textbf{r},t) \equiv \int_{0}^{2 \pi} d \varphi \textbf{e} \psi_{1} (\textbf{r},\varphi,t)
 \end{equation}
  and the nematic  tensor 
  \begin{equation}
   \textbf{Q} \equiv \int_{0}^{2 \pi} d \varphi (\textbf{e} \otimes \textbf{e} -  \mathbb{1} /2 )\psi_{1} (\textbf{r},\varphi,t) 
  \end{equation} 
Integrating the (mean-field) closed evolution equation of the one-body probability distribution, \cref{modeldescrip_eq:14}, we obtain the effective hydrodynamic equations for each one of the three fields 

\begin{equation}
\begin{split}
 \partial_{t} \rho (\textbf{r},t) =- \nabla \cdot \Big[ v_{\bar{\rho}} \textbf{p} -\Da \nabla \rho \Big]
\end{split}
\label{modeldescrip_eq:15}
\end{equation}

\begin{equation}
\begin{split}
 \partial_{t} \textbf{p} (\textbf{r},t) &= -\nabla \cdot \Big[v_{\bar{\rho}} (\frac{1}{2} \rho \mathbb{1} + \textbf{Q} )- \Da \nabla \textbf{p} \Big] \\
& \qquad  - \bar{\rho} \varepsilon \textbf{p}^{\perp} - D_{r} \textbf{p} 
 \end{split}
\label{modeldescrip_eq:16}
\end{equation}

\begin{equation}
\begin{split}
 \partial_{t} \textbf{Q} (\textbf{r},t) &= -\nabla \cdot \left(v_{\bar{\rho}}  \textbf{Y} \right)  -\frac{1}{4}\nabla \otimes(v_{\bar{\rho}} \textbf{p}) + \frac{1}{4}\nabla^{\perp} \otimes(v_{\bar{\rho}} \textbf{p}^{\perp})  \\
 &  \qquad + \Da \nabla^2 \textbf{Q}  - 2\bar{\rho} \varepsilon \textbf{Q}^{\perp}- 2D_{r} \textbf{Q} 
 \end{split}
\label{modeldescrip_eq:17}
\end{equation}
where $\perp$ indicates a rotation corresponding to $\bold{p}^{\perp}=\mathcal{R}\bold{p}$, $\nabla^{\perp}=\mathcal{R}\nabla$ and $\bold{Q}^{\perp}=\mathcal{R}\bold{Q}$ with  $\mathcal{R} = \left(\begin{array}{ccc} 0 & -1\\1 & 0\\ \end{array}\right)$.  

Note that the time evolution equation of each moment is linearly coupled to the next order moment. Consequently, the time evolution of the nematic tensor, \cref{modeldescrip_eq:17}, is coupled to the tensor $Y_{\alpha \beta \gamma} \equiv \int_{0}^{2 \pi} d \varphi (e_{\alpha} e_{\beta} e_{\gamma} -
\frac{1}{4} (\delta_{\alpha \beta} p_{\gamma} + \delta_{\alpha \gamma} p_{\beta} +\delta_{\beta \gamma} p_{\alpha} ))\psi_{1} (\textbf{r},\varphi,t)$, corresponding to the third moment of $\psi_{1}$.
As it is shown in the next subsection, $\rho (\textbf{r},t)$ constitutes the slowest moment of the probability distribution. Therefore, higher order moments are enslaved to $\rho (\textbf{r},t)$ and become irrelevant for the study of an infinite wave-length instability \cite{Dijkstra2019,Cates_runtumble2013}. This remark thus justifies to cut the hierarchy of equations by dropping the dependency on $Y_{\alpha \beta \gamma}$. As a result, we 
obtain a closed set of hydrodynamic equations accounting for the time evolution of the particle density, polarization vector and nematic field tensor.
Note that the hydrodynamic description we obtained is general and applies to any kind of interaction potentials $\mathcal{U}$ and $\mathcal{H}$. The different behaviors observed for ferromagnetic and nematic alignment have thus to be captured by the coefficients $\zeta$ and $\varepsilon$, which contain all the information regarding the microscopic interactions at this level of description.

\paragraph{Linear stability analysis} 
We now assume that the density $\rho(\textbf{r},t)$ is a slowly varying field. This is justified as long as  we are interested in the stability  of a homogeneous state. In this case, one can thus replace $\bar{\rho}$ by the local density field $\rho(\textbf{r},t)$ in the hydrodynamic equations \cite{Bialke2013}, where now 
\begin{equation}
v_{\bar{\rho}}\,\to\, v[\rho] = v_0 -\rho (\textbf{r},t) \zeta 
\end{equation}
We now perform a  linear stability analysis of the homogeneous and isotropic solution  $(\rho, \textbf{p} , \textbf{Q} )=(\overline{\rho}, 0,0)$ of the hydrodynamic  equations obtained after such replacement. To this end, we introduce a small perturbation, $\rho (\textbf{r}) = \bar{\rho} + \delta \rho$, $\textbf{p} (\textbf{r}) =  \delta \textbf{p}$ and $\textbf{Q} (\textbf{r}) = \delta \textbf{Q}$
and obtain a set of five independent linearized equations in Fourier space (see \cref{modeldescrip_eq:18,modeldescrip_eq:19,modeldescrip_eq:20} in \cref{appx_fourier_hydroeq} for details). 

Let us first focus on the linear stability in the absence of alignment interactions. As shown in \cite{Bialke2013}, considering the density and polarization fields only, the homogeneous solution in this case suffers a long wave-length instability associated to MIPS in a given parameter regime (see \cref{linear_stab_isotropic_disks} for full derivation). 
Adding the nematic field equation does not change this scenario, since \cref{modeldescrip_eq:17} has also a term proportional to the characteristic frequency $D_r$, ensuring a fast decay. It is then justified to assume that the fast moments, $\hat{p}_{\alpha}$ and $\hat{Q}_{\alpha \beta}$ (where  $\hat{*}$ denotes the Fourier transform), are adiabatically enslaved to the slow moment $\hat{\rho}$, associated to a conserved field (Goldstone theorem). 
This also justifies cutting the hierarchy of hydrodynamic equations, \cref{modeldescrip_eq:15,modeldescrip_eq:16,modeldescrip_eq:17} on the next (third) order moment $Y_{\alpha \beta \gamma}$, since it also relaxes faster than the density field. 

We then perform an adiabatic approximation on the polarization and nematic field, i.e. $ \partial_{t}p_{\alpha} \approx 0$ and $ \partial_{t} Q_{\alpha \beta} \approx 0 $, to obtain an effective diffusion equation in Fourier space,
\begin{equation}
\partial_{t} \delta \hat{ \rho}(\textbf{q})  =  \Da^{eff}_{\textbf{q}} \textbf{q}^2 \delta \hat{ \rho}(\textbf{q})  
\label{modeldescrip_eq:21}
\end{equation}
where
\begin{equation}
 \Da^{eff}_{\textbf{q}}=\left( \frac{1}{2} \left(v_0 - \bar{\rho} \zeta) (v_0 -2 \bar{\rho}  \zeta \right) \frac{A_{\textbf{q}}}{A_{\textbf{q}}^2 + B_{\textbf{q}}^2} - \Da  \right)
 \end{equation}
with
\begin{align}
&A_{\textbf{q}} = -\left(\left(\frac{1}{4}\frac{\left(v_0 - \bar{\rho} \zeta \right)^2}{\Da \textbf{q}^2 + 2 D_r} \frac{1}{1+ (\frac{2 \bar{\rho} \varepsilon}{\Da \textbf{q}^2 + 2 D_r})^2} + \Da \right) \textbf{q}^2 + D_r\right) \\
&B_{\textbf{q}} = -\left(\frac{1}{4}\frac{\left(v_0 - \bar{\rho} \zeta \right)^2}{\left(\Da \textbf{q}^2 + 2 D_r \right)^2}  \frac{2 \bar{\rho} \varepsilon}{1+ (\frac{2 \bar{\rho} \varepsilon}{\Da \textbf{q}^2 + 2 D_r})^2} \textbf{q}^2 -\bar{\rho} \varepsilon \right)
\label{modeldescrip_eq:22}
\end{align}

A linear instability at a wave vector $\textbf{q}$ is signaled by a negative effective diffusion coefficient  $ \Da^{eff}_{\textbf{q}} < 0$. The limit of linear stability of the homogenous disordered gas can therefore be computed by setting $ \Da^{eff} = 0$, which at $\textbf{q} \rightarrow 0$ leads to
\begin{equation}
\begin{split}
8 \left(\frac{v_0}{v^*} - \tilde{\zeta} \right) \left(\frac{v_0}{v^*} - 2 \tilde{\zeta}  \right) \frac{1}{1 + \tilde{\varepsilon}^2} + 1 = 0 \, .
\end{split}
\label{modeldescrip_eq:22}
\end{equation}
Here, $\frac{v_0}{v^*}$ is the reduced self-propulsion speed, where $v^*=4\sqrt{\Da D_r}$. The translational and rotational friction coefficients are also written in their dimensionless form, $\tilde{\zeta} =\frac{\bar{\rho}}{v^*}\zeta$ and $\tilde{\varepsilon} =\frac{\bar{\rho}}{D_r}\varepsilon$, and encode all the specificities of the (anisotropic) interactions between particles.  \cref{modeldescrip_eq:22} thus constitutes our mean-field prediction of the spinodal of a system of self-propelled disks subjected to generic aligning torques, explicitly written in terms of the relevant non-dimensional parameters of the model $(\frac{v_0}{v^*},\tilde{\zeta},\tilde{\varepsilon} )$.

\section{Continuum theory vs. microscopic simulations} \label{comparison}

In this section, we confront quantitatively the prediction from the linear stability of the hydrodynamic equations with direct numerical simulations of ABP with different alignment interactions. To do so, we will employ the $\mathrm{Pe}$ as the control parameter quantifying the degree of activity in the system, instead of $v_0/v^*$ , as it is customary in simulations of ABP. Both parameters are related through 
\begin{equation} 
v_0/v^* = \frac{R}{4\sqrt{\Da/D_r}}\mathrm{Pe} = 0.638\mathrm{Pe}
\end{equation} 

where we have numerically computed the value of $\Da$, see \cref{compu_param}, taken from the long-time diffusion coefficient of a passive system ($v_0 = 0$).

We have derived the evolution equation for the one-body probability distribution $\psi_1$, where the two-body correlations have been cast into two effective friction coefficients $\tilde{\zeta}$ and $\tilde{\varepsilon}$ which are given by integrals of the the pair correlation function $\mathcal{G}(r,\theta,\varphi,t)$ (we only consider steady-states and therefore, from now on, we drop the time dependency). The excluded volume potential  imposes a planar rotational symmetry, $ \mathcal{G}(r,\theta,\varphi) = \mathcal{G}(r,-\theta,\varphi) $, and the alignment ones (both ferromagnetic and nematic) impose $ \mathcal{G}(r,\theta,\varphi) = \mathcal{G}(r,\theta,-\varphi) $.  As a result, $\tilde{\varepsilon}=0$ and the limit of stability thus reduces to 
\begin{equation}
8 \left(\frac{v_0}{v^*} - \tilde{\zeta} \right) \left(\frac{v_0}{v^*} - 2 \tilde{\zeta}  \right)  + 1 = 0 
\end{equation}

In order to test this prediction against  particle-based simulations, we now have to compute $\tilde{\zeta}$ for different values of the microscopic parameters and determine whether they fall or not in the instability region predicted by the hydrodynamic model. 
As  $\tilde{\zeta}$ is given by the pair correlation function, we first start by assessing the impact that the different interactions have on the in-plane structure of the system, defined by coordinates $r$ and $\theta$, see \cref{continuum-model_fig:1}. 

To this end, we compute $G(r,\theta) = \int_{0}^{2\pi} d\varphi \:\mathcal{G}(r,\theta,\varphi)$ using Brownian dynamics simulations.
In \cref{compar_meanfield_microsc_gr_fig:1} we show $G(r,\theta=0)$ and $G(r,\theta=\pi)$, the pair correlation function ahead and behind a tagged particle, for four representative cases: (a) isotropic passive particles, (b) isotropic ABP, (c) ferromagnetic ABP and (d) nematic ABP; using fixed Pe=16.6 and $g/g_{f,n}=0.4$ (below the onset of orientational order) in the presence of  alignment. [In \cref{compar_meanfield_microsc_gr_appx_fig:1} (a) - (d) in \cref{correl_funct_appx} we show the full $G(r,\theta)$.]
As expected, $G(r,\theta)$ is isotropic for a passive suspension of disks, thus yielding $\zeta=0$. 
Activity breaks this spatial isotropy: it is more likely for the tagged particle to find other particles in front of it ($\theta=0$)  than behind ($\theta=\pi$), giving rise to the self-trapping mechanism at the origin of MIPS. Particles block each other in the direction of self-propulsion, giving rise to a reduction of $v_{\bar{\rho}}$. Such anisotropy of $G(r,\theta)$ is at the origin of the non-zero value of $\zeta$, which quantifies the decay rate of the self-propulsion speed with the density. 

In the presence of ferromagnetic alignment, the structural anisotropy in the system is enhanced as compared to the non-aligning ABP case. The peak structure of $G(r,\theta)$ is more pronounced, as also is the contrast between $G(r,\theta=0)$ and $G(r,\theta=\pi)$. Thus, the mutual kinetic arrest due to collisions just described is also enhanced and as a result also the aggregation of particles. This is in qualitative agreement with our earlier results Fig. \ref{phasediag_fig:1}, showing  that MIPS occurs at lower values of Pe and $\phi$ as the ferromagnetic coupling is increased. The peak structure of $G(r,\theta)$ at short distances in this case also shows that ferromagnetic ABP aggregate into denser structures with a higher degree of spatial order than their isotropic counterpart at a given Pe. 

The spatial distribution of particles subjected to nematic alignment seems not to be significantly affected by the presence of torques. In   \cref{compar_meanfield_microsc_gr_fig:1} (d) we show $G(r,\theta=\pi)$ and $G(r,\theta=0)$ for a nematic coupling of the same relative strength as the one used for the ferromagnetic case.  The results show that nematic alignment does not significantly affect the self-trapping phenomenon triggered by the competition between self-propulsion and excluded volume interactions. 

Overall, the interpretation of the $G(r,\theta)$ from the viewpoint of the microscopic theory leading to the hydrodynamic equations is in qualitative agreement with the results presented in \cref{phasediag_fig:1} and \cref{binodal}. The emergence of MIPS can be understood as particles blocking each other preferentially along their self-propulsion direction, a mechanism well captured by the appearance of anisotropy in the pair correlation functions. Ferromagnetic interactions favor MIPS, while nematic ones do not significantly affect it. In the following, we push this picture further, and show that it can provide a quantitive agreement with particle-based simulations through the calculation of $\zeta$.

 \begin{figure}
	\centering
	\includegraphics[trim=100 0 200 0,width=\columnwidth]{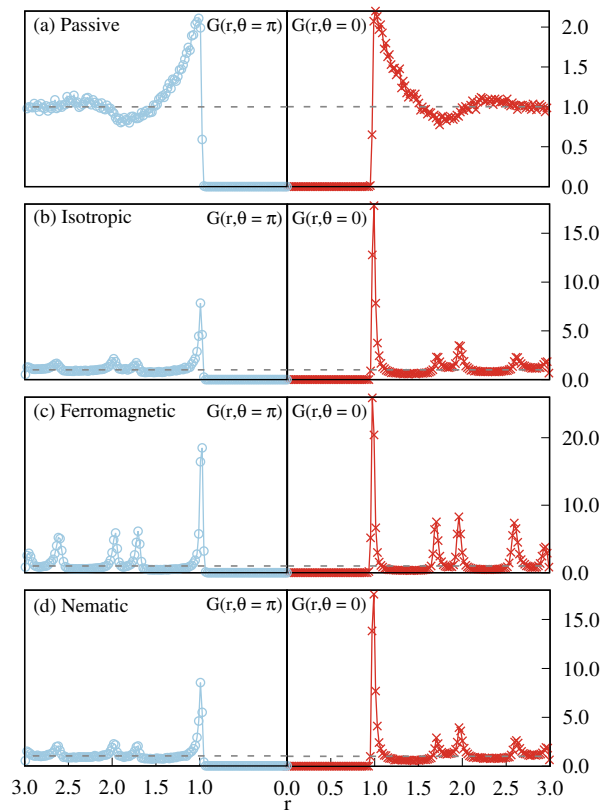}
	\caption{Radial distribution function along two opposite directions of the ($r$, $\theta$)-plane, corresponding to the front ($\theta=0$, right column) and the back ($\theta=\pi$, left column) of the tagged particle  for (a) $\mathrm{Pe} = 0$ and $g=0$ (passive suspension); (b) $\mathrm{Pe} = 16.6$ and $g=0$; (c) $\mathrm{Pe}= 16.6$ and ferromagnetic $g/g_f = 0.4$; (d) $\mathrm{Pe} =  16.6$ and nematic $g/g_n = 0.4$. }
	\label{compar_meanfield_microsc_gr_fig:1}
	\end{figure}

The limit of stability of the homogeneous and isotropic phase predicted by the hydrodynamic model in dimensionless units is,
\begin{equation}
\begin{split}
\tilde{\zeta}^{\pm} = \frac{3}{4}\frac{v_0}{v^*} \pm \frac{1}{4} \sqrt{\left(\frac{v_0}{v^*}\right)^2 - 1 }
\end{split}
\label{modeldescrip_eq:23}
\end{equation}
In all the region encompassed between $\tilde{\zeta}^{-} < \tilde{\zeta} < \tilde{\zeta}^{+}$ the homogeneous state is unstable. This unstable region is represented  in blue in the $(\mathrm{Pe}, \tilde{\zeta})$ diagrams, \cref{onsetphasesep_fig:2} (a) and (b). 
In the same figures we also plot  $\tilde{\zeta}$, computed from the $\mathcal{G}(r,\theta,\varphi)$ obtained from Brownian dynamics simulations. 

Let us first focus on the numerical values of $\tilde{\zeta}$ computed at $g=0$ and depicted in  \cref{onsetphasesep_fig:2} as a function of Pe. 
As Pe grows, $\tilde{\zeta}$ also does, as a consequence of a larger anisotropy in the pair correlation function for increasing activity. Beyond Pe$\approx16$, the value for which $\tilde{\zeta}$ eventually penetrates the instability region, the rate of growth of $\tilde{\zeta}$ increases significantly. Such behavior is due to the presence of  a finite fraction of slow particles belonging to a dense cluster. At large Pe, the slow-down of particles due to the collision persistence is enhanced, triggering the feed-back mechanism by which the phase separation takes place. The value of Pe at which $\tilde{\zeta}$ crosses $\tilde{\zeta}^-$ is identified with the onset of MIPS.   

\begin{figure}
	\centering
	 \includegraphics[trim=270 20 70 16, width=\columnwidth]{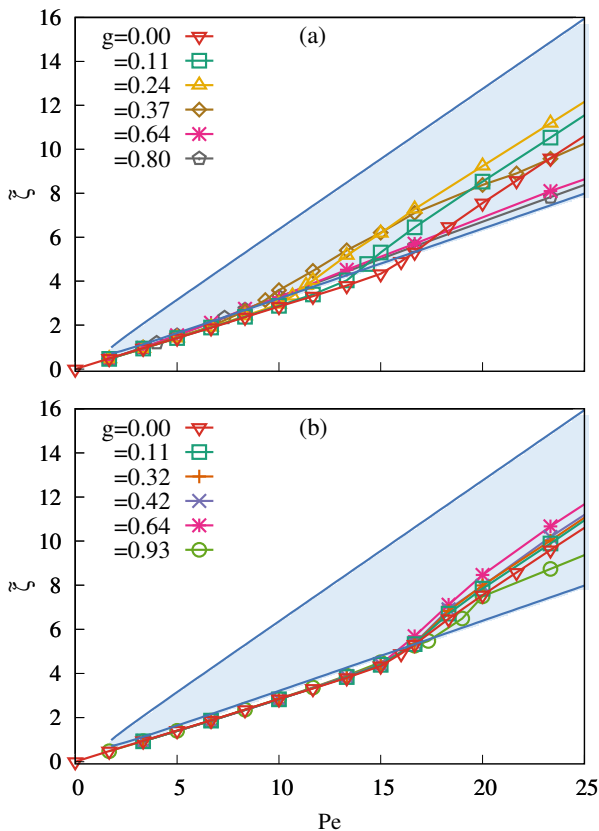}
	\caption{Linear instability region at $\tilde{\varepsilon} = 0$ (blue region) together with the $\tilde{\zeta}$ coefficient computed from particle-based simulations as a function of the $\mathrm{Pe}$ for different values of (a) ferromangetic and (b) nematic alignment strength.}
	\label{onsetphasesep_fig:2}
	\end{figure}

We turn now our attention to the system with ferromagnetic alignment. The values of $\tilde{\zeta}$ extracted from simulations are shown in \cref{onsetphasesep_fig:2} (a).
As we already pointed out, the anisotropy in the correlation function $\mathcal{G}(r,\theta,\varphi)$ is enhanced by the presence of ferromagnetic alignment. As a result, the numerical values of $\tilde{\zeta}$ are now larger than the ones in the $g=0$ case. The penetration into the instability region happens at lower values of $\mathrm{Pe}$. The values of Pe at which the curves of $\tilde{\zeta}$ enter the instability region are indicated in \cref{phasediag_fig:1} (a), showing that the prediction of the microscopic theory matches the numerical simulation results accurately.

In the case of nematic alignment, the structural anisotropy of the system is not enhanced as compared to the isotropic $g=0$ case. Thus, the change in tendency of the numerical values of $\tilde{\zeta}$ and the crossing of the limit of stability happens at approximately the same value of $\mathrm{Pe}$, regardless of the alignment interaction strength, see \cref{onsetphasesep_fig:2} (b). This prediction also agrees well with the numerical simulation results,  \cref{phasediag_fig:1} (b).

In our Brownian dynamics simulations, we observed that the region of coexistence is shifted to lower values of $\mathrm{Pe}$ as the ferromagnetic coupling is increased, \cref{binodal} (a). On the contrary, for nematic alignment, the coexistence region remains at the same values of $\mathrm{Pe}$ within our numerical accuracy,  see \cref{binodal} (b). 
We can therefore conclude that the results obtained from particle-based simulations and the continuum model are in good agreement, and predict the same phase behavior: ferromagnetic torques enhance the aggregation of particles and the formation of MIPS, while nematic torques have a neutral effect. 
 
\section{Conclusions}
We have introduced a 2D model of spherical Active Brownian Particles, subjected to both excluded volume  and velocity-alignment interactions. Our model decouples the alignment mechanism from steric effects, allowing to disentangle these two and tune the strength of nematic or ferromagnetic alignment with no need of introducing shape anisotropy.  

We studied such model system both analytically and numerically. First, in order to grasp the role played by the self-propulsion and alignment strength, we explored the phase behavior of the system by varying the Pe and coupling strength $g$ by means of Brownian dynamics simulations.  We identified  the emergence of oriented states  featuring different collectively moving structures, such as polar and nematic lanes. Unlike systems of self-propelled elongated particles, here the nematic phase keeps the  symmetry of the interaction, with no signature of large-scale polar lanes.

We then focused on the main aspect of the present work, namely phase separation, or MIPS, triggered by the competition between excluded volume effects and self-propulsion in the presence of velocity alignment, yet in the absence of global orientational order. We systematically derived a continuum description of the system taking as a starting point the $N$-body Smoluchowski equation. This  yields a set of coupled hydrodynamic equations for the density, polarization and nematic  fields, which directly follow from the microscopic Langevin equations. A key advantage of this approach, which generalizes the work in \cite{Bialke2013}  to include alignment interactions, is that the resulting field equations are written in terms of the microscopic parameters  of the particle-based model, allowing for an explicit comparison between the two, as opposed to more phenomenological approaches based on symmetry arguments \cite{toner2005hydrodynamics, CatesField}. 

The microscopic mechanism giving rise to MIPS within the theory, is a long-wave length linear instability of the homogenous disordered state due to the anisotropy of the pair distribution function. The latter arises from the fact that particles have a higher tendency to collide with their neighbors along the direction of self-propulsion, leading to the formation of clusters and eventually a full phase separation. We have shown that such self-trapping mechanism is enhanced in the presence of ferromagnetic alignment but remains largely unaffected in the nematic case. The predictions of the onset of MIPS from the microscopic theory agree quantitatively with particle-based simulations. 

We showed that the mechanism behind MIPS appears to be preserved in the presence of alignment interactions. The present study opens the possibility of extending the this formalism to study more complex situations. For instance, studying how chirality (circle swimming) \cite{BennoPRL, BennoJCM, Sabine2018, ma2021dynamical}, quench disorder \cite{reichhardt2014active,bhattacharjee2019confinement,chardac2021emergence} or other kind of interactions \cite{matas2014hydrodynamic,yoshinaga2017hydrodynamic, liao2020dynamical} affect MIPS, constitute interesting lines of future investigation. 
Here, we have focused on the long-wave length instability of the disordered state. However, other linear instabilities can take place. This means that, in different parameter regimes, the theory might be able to account for the richness of different structures observed in particle-based simulations across the phase diagram. As a next step, it would be very interesting to extend the theory and explore all the unstable modes associated to both the homogeneous disordered state and the homogeneous oriented one. We leave this challenging task for future work.  

\section{Acknowledgments}
E. S.-S. thanks Joan Codina for helpful discussions. D. L. acknowledges Ministerio de Ciencia, Innovaci\'on y Universidades MCIU/AEI/FEDER for financial support under grant agreement RTI2018-099032-J-I00.
I. P. acknowledges support from Ministerio de Ciencia, Innovaci\'on y Universidades MCIU/AEI/FEDER for financial support under grant agreement PGC2018-098373-B-100 AEI/FEDER-EU and from Generalitat de Catalunya under project 2017SGR-884. E. S.-S. and I. P. acknowledge Swiss National Science Foundation Project No. 200021-175719.

\newpage

\appendix

\section{Derivation of the nematic torque}\label{deriv_nemat_hamilt}

We write the Hamiltonian of a system of N particles subjected to a local nematic alignment interaction,
 \begin{equation}
\Ha =- J \sum_i \sum_{j \in \omega_i} \textbf{q}_i \cdot \textbf{q}_j
\label{appx_hamilt-nemat_eq:1}
\end{equation}
where the uniaxial nematic tensor reads $ \textbf{q}_i = \textbf{e}_i \otimes \textbf{e}_i - \frac{1}{2} \mathbb{1}$ and $\textbf{e}_{i} = \left(\cos \varphi_i,\sin \varphi_i \right)$ and $\textbf{a} \cdot \textbf{b} = \Tr(\textbf{a} \textbf{b})$. In the case of study, the trace can be expressed as,

 \begin{equation}
\Tr(\textbf{q}_i \textbf{q}_j) = \frac{1}{2} \left( \cos 2 \varphi_i   \cos 2 \varphi_j +\sin 2\varphi_i \sin 2\varphi_j  \right)
\label{appx_hamilt-nemat_eq:2}
\end{equation}

Using the trigonometric relation $~\cos \left( \varphi_{i}-\varphi_{j} \right) = \cos  \varphi_i  \cos  \varphi_j + \sin  \varphi_i  \sin  \varphi_j~$, it is possible to rewrite the Hamiltonian in terms of the phase difference $\varphi_{ij} =  \varphi_i  - \varphi_j $, 

\begin{equation}
\Ha = - J \sum_i \sum_{j \in \omega_i} \frac{1}{2} \cos \left( 2 \varphi_{ij} \right)
\label{appx_hamilt-nemat_eq:3}
\end{equation}


Further trigonometric relations can be used to write the Hamiltonian in different ways, such as $\cos  \left(2 \varphi_{ij}  \right) = 2\cos^2 \left( \varphi_{ij}  \right) -1 $, which leads to,
\begin{equation}
\Ha = - J \sum_i \sum_{j \in \omega_i}\left(  \cos^2  \varphi_{ij}    - \frac{1}{2} \right)
\label{hamilt-nemat_eq:4}
\end{equation}
Deriving with respect to an angle on gets the desired expression for the torque, 
\begin{equation}
T_i = -\frac{\partial \Ha }{\partial \varphi_i} = -  J \sum_{j}   \sin \left( 2 \varphi_{ij} \right) 
\label{hamilt-nemat_eq:4}
\end{equation}


\section{Characterization of MIPS} \label{appx_phase-sep-characteriz}

 We compute the fraction of particles in the largest cluster of the system, $\Pi$, \cref{appx_onsetphasesep_fig:1}. 
 Clusters are defined by setting a threshold distance below which two particles are considered to belong to the same cluster. Here, we set this threshold to be the cutoff distance of the steric potential, $R=1$, and we compute $\Pi$ as a function of $\mathrm{Pe}$. As $\mathrm{Pe}$ increases, collisions between self-propelled particles become more probable, thus enhancing mutual blocking due to the swimming persistence. This leads to further aggregation between particles and, consequently, to the growth of $\Pi$.
  
 \begin{figure}[h]
	\centering
	\includegraphics[trim=100 50 0 220, width=\columnwidth]{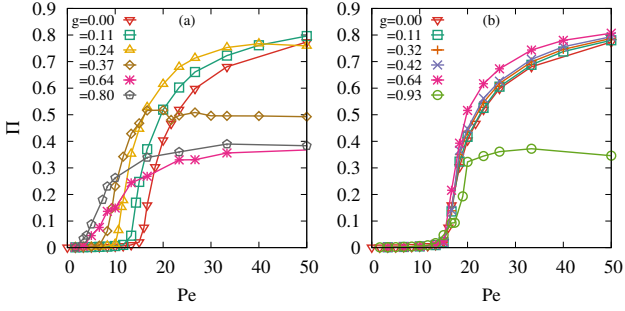}
	\caption{Probability to belong to the largest cluster as a function of the normalized swimming speed $\mathrm{Pe}$ at different values of the (a) ferromagnetic and (b) nematic alignment strengths.}
	\label{appx_onsetphasesep_fig:1}
	\end{figure}

Nevertheless, the curves of $\Pi(\mathrm{Pe})$ do not tend to $\Pi=1$ as $\mathrm{Pe}$ is increased, but they saturate at lower values of $\Pi$, when $g$ approaches the flocking phase transition from below. This is due to the fact that local orientational correlations grow at $g \neq 0$, setting a different characteristic interparticle length than the one induced by purely excluded volume interactions. As a result, particles are further apart and clusters defined according to the cutoff distance $R=1$ are now smaller in size.

\section{Finite size effects}\label{system_size}

We plot in \cref{appx_systemsize_fig:1} the binodals at a fixed nematic alignment strength $g =0.64$ and for three different system sizes $N=4000, 8000, 16000$.
\begin{figure}[h]
	\centering
	 \includegraphics[trim=0 0 0 0, width=\columnwidth]{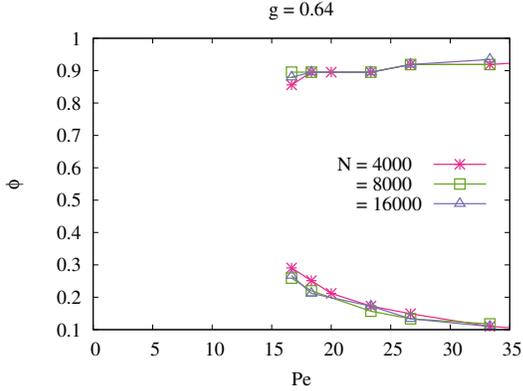}
	\caption{Phase coexistence regions for fixed nematic alignment strength $g =0.64$ and different system sizes $N=4000, 8000, 16000$. }
	\label{appx_systemsize_fig:1}
	\end{figure}
	
In the same spirit, we also compute the numerical values of $\tilde{\zeta}$ at fixed nematic alignment strength $g =0.64$ for system sizes $N=8000, 16000$, \cref{appx_systemsize_fig:2}. We observe that the penetration of $\tilde{\zeta}$ into the instability region occurs at the same value of $\mathrm{Pe}$ as it does for $N=4000$, within numerical accuracy. 
	
\begin{figure}[h]
	\centering
	 \includegraphics[trim=0 0 0 0, width=\columnwidth]{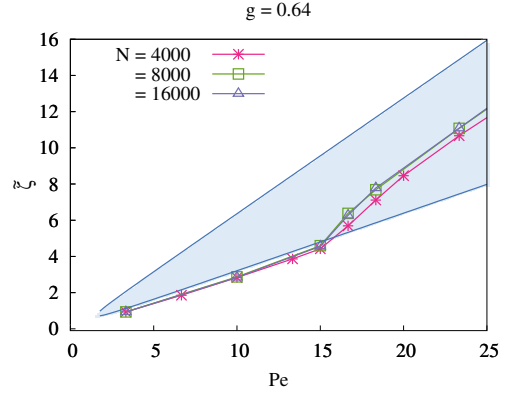}
	\caption{Linear instability region at $\tilde{\varepsilon} = 0$ (blue region) together with the $\tilde{\zeta}$ coefficient computed from particle-based simulations as a function of the $\mathrm{Pe}$ for a fixed nematic alignment strength $g =0.64$ and different system sizes $N=4000, 8000, 16000$.}
	\label{appx_systemsize_fig:2}
	\end{figure}

We can therefore state that the numerical data obtained at $N=8000, 16000$ agree well with the one obtained at $N=4000$, showing the robustness of the results presented to characterize the phase separation induced by the motility.

\section{Continuum model} \label{appx_continuum_model_derivation}
\subsection{Expressions of force and torque} \label{appx_continuum_model_forcetorque}

The mean force and torque exerted by the surrounding particles into the tagged particle (labeled \textit{1}) are expressed as,

\begin{equation}
\begin{split}
&\textbf{F} (\textbf{r}_{1},\varphi_{1};t)=\int_{-\infty}^{\infty} d\textbf{r}_{2} \int_{0}^{2\pi} d\varphi_{2} u ' \left(r_{12}\right)\frac{\textbf{r}_{12}}{r_{12}} \psi _{2}
\end{split}
\label{appx_continuum_model_forcetorque_eq:1}
\end{equation}
and the mean torque is, 
\begin{equation}
\begin{split}
T(\textbf{r}_{1},\varphi_{1};t) = \int_{-\infty}^{\infty} d\textbf{r}_{2}  \int_{0}^{2\pi} d\varphi_{2} v\left(r_{12}\right) w' \left( \varphi_{12} \right) \psi _{2}
\end{split}
\label{appx_continuum_model_forcetorque_eq:2}
\end{equation}

Introducing the decomposition of $\psi_{2}(\textbf{r}_1,\textbf{r}_2,\varphi_{1},\varphi_{2},t)$ stated in \cref{theoret-mean-field_eq:5}, as well as the changes of variables $\textbf{r}_{12} = \textbf{r}_{2} - \textbf{r}_{1}$ and  $\varphi_{12} = \varphi_{2} - \varphi_{1}$, which yield $ d  \textbf{r}_{12} =d\textbf{r}_{2} $ and $ d  \varphi_{12} =d\varphi_{2} $, it is possible to rewrite \cref{appx_continuum_model_forcetorque_eq:1,appx_continuum_model_forcetorque_eq:2} as,

\begin{equation}
\begin{split}
\textbf{F}\left(\textbf{r}_{1},\varphi_{1};t\right)&=\bar{\rho} \psi_{1}\left(\textbf{r}_1,\varphi_{1},t\right)\int_{-\infty}^{\infty} d\textbf{r}_{12} \int_{0}^{2\pi}d\varphi_{12}\\
& \qquad u ' \left(r_{12} \right) \frac{\textbf{r}_{12}}{r_{12}}  \mathcal{G}\left(r_{12},\theta,\varphi_{12},t\right)
\label{appx_continuum_model_forcetorque_eq:3}
\end{split}
\end{equation}
\begin{equation}
\begin{split}
T \left(\textbf{r}_{1},\varphi_{1};t\right) &=  \bar{\rho} \psi_{1}\left(\textbf{r}_1,\varphi_{1},t\right) \int_{-\infty}^{\infty} d\textbf{r}_{12} \int_{0}^{2\pi} d\varphi_{12}\\
 &\qquad v\left(r_{12} \right)    w '\left( \varphi_{12}\right)  \mathcal{G}\left(r_{12},\theta,\varphi_{12},t\right)
\label{appx_continuum_model_forcetorque_eq:4}
\end{split}
\end{equation}

It is straightforward to see that the projection of the force in the direction of self-propulsion, $\textbf{e}$, can be expressed as $\textbf{e} \cdot \textbf{F} = -\bar{\rho} \psi_{1} \zeta$, where $\zeta$ has the functional form \cref{theoret-mean-field_eq:7}. 
Similarly, the torque can also be decomposed in $ T = -\bar{\rho} \psi_{1} \varepsilon$, where $\varepsilon$ is expressed as in \cref{theoret-mean-field_eq:9}.

\subsection{Gram-Schmidt orthonormalization of the force} \label{appx_GSortho}

We perform a Gram-Schmidt orthonormalization on the force $\textbf{F} $ to decompose it in the vector basis formed by the direction of self-propulsion and the gradient of the probability distribution, $\{\textbf{e},\nabla \psi_{1}\}$. 

We pick the first vector of the orthonormal set $\{ \textbf{u}_1, \textbf{u}_2\}$ we want to construct, $\textbf{u}_1 = \textbf{e}$. This one already fulfils $|\textbf{e}| = 1$. Then the second vector is, 

\begin{equation}
\begin{split} 
 \textbf{u}_2 = \nabla_1 \psi_1 - \mathrm{proj}_{ \textbf{u}_1}\left( \nabla_1 \psi_1 \right)
\end{split}
\label{appx_gram_schmidt-eq:1}
\end{equation}

where the projection operator is $\mathrm{proj}_{ \textbf{a}}\left( \textbf{b} \right) = \frac{\textbf{a} \cdot \textbf{b}}{\textbf{a} \cdot \textbf{a}}\textbf{a}  $, giving the projection of vector $ \textbf{b}$ along the axis spanned by  \textbf{a}. Thus, we can rewrite \cref{appx_gram_schmidt-eq:1},

\begin{equation}
\begin{split} 
 \textbf{u}_2 = \nabla_1 \psi_1 - \left( \textbf{e} \cdot \nabla_1 \psi_1 \right) \textbf{e}
\end{split}
\label{appx_gram_schmidt-eq:2}
\end{equation}

We now proceed to normalize $ \textbf{u}_2$,

\begin{equation}
\begin{split} 
\textbf{u}_2 = \frac{\nabla_1 \psi_1 - \left( \textbf{e} \cdot \nabla_1 \psi_1 \right) \textbf{e}}{|\nabla_1 \psi_1 - \left( \textbf{e} \cdot \nabla_1 \psi_1 \right) \textbf{e}|}
\end{split}
\label{appx_gram_schmidt-eq:3}
\end{equation}

We have thus constructed an orthonormal vector basis. We can now decompose the force, 

\begin{equation}
\begin{split} 
 \textbf{F}& = \left( \textbf{e} \cdot  \textbf{F} \right)\textbf{e} \\
 &+ \left(  \frac{\nabla_1 \psi_1 - \left( \textbf{e} \cdot \nabla_1 \psi_1 \right) \textbf{e}}{|\nabla_1 \psi_1 - \left( \textbf{e} \cdot \nabla_1 \psi_1 \right) \textbf{e}|} \cdot   \textbf{F} \right) \frac{\nabla_1 \psi_1 - \left( \textbf{e} \cdot \nabla_1 \psi_1 \right) \textbf{e}}{|\nabla_1 \psi_1 - \left( \textbf{e} \cdot \nabla_1 \psi_1 \right) \textbf{e}|} \\
 &= \Big[\left( \textbf{e} \cdot  \textbf{F} \right) \Big.\\
 &\qquad\Big. - \left(  \frac{\left(\nabla_1 \psi_1 - \left( \textbf{e} \cdot \nabla_1 \psi_1 \right) \textbf{e} \right)\cdot   \textbf{F} }{|\nabla_1 \psi_1 - \left( \textbf{e} \cdot \nabla_1 \psi_1 \right) \textbf{e}|^2} \right) \left( \textbf{e} \cdot \nabla_1 \psi_1 \right)\Big]\textbf{e} \\
&\qquad\qquad+ \left(  \frac{\left(\nabla_1 \psi_1 - \left( \textbf{e} \cdot \nabla_1 \psi_1 \right) \textbf{e} \right)\cdot   \textbf{F} }{|\nabla_1 \psi_1 - \left( \textbf{e} \cdot \nabla_1 \psi_1 \right) \textbf{e}|^2} \right)\nabla_1 \psi_1 
\end{split}
\label{appx_gram_schmidt-eq:4}
\end{equation}

We first consider that the projection of the force along  $\frac{\nabla_1 \psi_1 - \left( \textbf{e} \cdot \nabla_1 \psi_1 \right) \textbf{e}}{|\nabla_1 \psi_1 - \left( \textbf{e} \cdot \nabla_1 \psi_1 \right) \textbf{e}|}$ is much smaller than its projection along $\textbf{e}$. Second, we also consider that $|\nabla_1 \psi_1 - \left( \textbf{e} \cdot \nabla_1 \psi_1 \right) \textbf{e}|^2 \approx |\nabla_1 \psi_1 |^2$, assuming that $\textbf{e} $ and $\nabla_1 \psi_1$ are 'almost' perpendicular vectors. This leads to, 

\begin{equation}
\begin{split} 
 \textbf{F} & \approx  \left( \textbf{e} \cdot  \textbf{F} \right) \textbf{e}+ \left(  \frac{\left(\nabla_1 \psi_1 - \left( \textbf{e} \cdot \nabla_1 \psi_1 \right) \textbf{e} \right)\cdot   \textbf{F} }{|\nabla_1 \psi_1|^2} \right)\nabla_1 \psi_1 \\
 &= \big(\textbf{e} \cdot \textbf{F} \big )\textbf{e} + \big(1-\Da \big)\nabla\psi_{1}
\end{split}
\label{appx_gram_schmidt-eq:4}
\end{equation}

where the first term on the right hand side (RHS)  corresponds to \cref{theoret-mean-field_eq:6} and $\Da$ corresponds to \cref{modeldescrip_eq:14bis}.

\subsection{Fourier transform of the hydrodynamic equation}\label{appx_fourier_hydroeq}

Writing  \cref{modeldescrip_eq:15,modeldescrip_eq:16,modeldescrip_eq:17} in Fourier space, 
 \begin{equation}
\textbf{u} \sim \hat{\textbf{u} }e^{i \textbf{q} \cdot \textbf{r}}
\label{modeldescrip_eq:18bis}
\end{equation}
where $\textbf{u}=\left(\delta \rho, \delta p_x,\delta p_y, \delta Q_{xx}, \delta Q_{xy}\right)$,
we obtain the following time evolution equations for the perturbation, 
 \begin{equation}
\partial_{t} \delta \hat{ \rho}  =- i q_{\beta} \left[ \left(v_0- \bar{\rho} \zeta \right) \delta\hat{p}_{\beta}-i q_{\beta} \Da \delta\hat{\rho}\right] 
\label{modeldescrip_eq:18}
\end{equation}
 \begin{equation}
 \begin{split}
\partial_{t} \delta \hat{p}_{\alpha} &= -  i q_{\beta} \left[\frac{1}{2} \left(v_0 -2  \bar{\rho}\zeta  \right)  \delta \hat{\rho} \delta_{\alpha \beta}+\left(v_0 - \bar{\rho} \zeta \right) \delta  \hat{Q}_{\alpha \beta} \right. \\
\qquad & -  \left.  i q_{\beta} \Da \delta \hat{p}_{\alpha} \right]  -  \bar{\rho} \varepsilon \delta\hat{p}^{\perp}_{\alpha}  - D_{r}\delta \hat{p }_{\alpha}
\end{split}
\label{modeldescrip_eq:19}
\end{equation}

\begin{equation}
\begin{split}
 \partial_{t} \delta \hat{Q}_{\alpha \beta}  &= -i q_{\gamma} \left[\left(v_0 - \bar{\rho} \zeta \right) \Big(- \frac{1}{2} \delta_{\alpha \beta} \delta \hat{p}_{\gamma} +\frac{1}{4} \left(\delta_{\alpha \beta} \delta \hat{p}_{\gamma} \right. \right.\\
  \qquad &+ \left. \left. \delta_{\alpha \gamma} \delta \hat{p}_{\beta} +\delta_{\beta \gamma} \delta \hat{p}_{\alpha} \right)\Big) - i q_{\gamma} \Da \delta \hat{Q}_{\alpha \beta}  \right]\\
  \qquad & - 2 \bar{\rho}  \varepsilon  \delta \hat{Q}_{\alpha \beta}^{\perp} - 2 D_{r} \delta \hat{Q}_{\alpha \beta}
 \end{split}
\label{modeldescrip_eq:20}
\end{equation}
where $q_{\beta} $ is the $\beta$-component of the wave vector. \Cref{modeldescrip_eq:18,modeldescrip_eq:20} constitute a system of 5 independent evolution equations (the nematic tensor has just two independent components due to the traceless and symmetric conditions). 

\subsection{Linear stability analysis of a system of isotropic repulsive disks}\label{linear_stab_isotropic_disks}

We consider a system of isotropic polar active disks without alignment interactions. 
Since there are no local torques ($\varepsilon= 0$) and the only effective interaction between particles is of steric origin, we cut the hierarchy of effective hydrodynamic equations to first order. Hence, the time evolution for the density and the polarization field in Fourier space now reads, 
 \begin{equation}
\partial_{t} \delta \hat{ \rho}  =- i q_{\beta} \Big[ (v_0- \bar{\rho} \zeta) \delta\hat{p}_{\beta}-i q_{\beta} \Da \delta\hat{\rho}\Big] 
\label{linear_stab_eq:1}
\end{equation}
 \begin{equation}
 \begin{split}
\partial_{t} \delta \hat{p}_{\alpha} &= -  i q_{\beta} \Big[\frac{1}{2} (v_0 -2  \bar{\rho} \zeta)  \delta \hat{\rho} \delta_{\alpha \beta} -  i q_{\beta} \Da \delta \hat{p}_{\alpha} \Big] - D_{r}\delta \hat{p }_{\alpha}
\end{split}
\label{linear_stab_eq:2}
\end{equation}

Writing the system of equations in matrix form $\partial_{t}(\delta \hat{\rho} \;   \delta \hat{p}_x  \; \delta \hat{p}_y )^{T} = M (\delta \hat{\rho} \;   \delta \hat{p}_x  \; \delta \hat{p}_y )^{T}$, where
\[
M=
 \begin{bmatrix}
  - \Da\textbf{q}^2 & - i (v_0 - \bar{\rho} \zeta) q_x & - i (v_0 - \bar{\rho} \zeta) q_y  \\
    - i \frac{1}{2} (v_0 -2\bar{\rho}  \zeta)q_x & - (\Da \textbf{q}^2+D_r ) & 0\\
    - i \frac{1}{2}(v_0 -2 \bar{\rho} \zeta) q_y & 0 &  - (\Da \textbf{q}^2+D_r )
  \end{bmatrix}
\]
it is possible to compute its eigenvalues $\det(M - \lambda \mathbb{1}) = 0$, which are, 
\begin{equation}
\begin{split}
\lambda_{1} &=-(D_r + \Da \textbf{q}^2 )\\
\lambda_{2}&=-\frac{1 }{2}(2 \Da \textbf{q}^2 +D_r)  + \frac{1 }{2}\sqrt{D_r^2 - 2\textbf{q}^2  (v_0 - \bar{\rho} \zeta) (v_0 -2 \bar{\rho} \zeta)}\\
\lambda_{3}&=-\frac{1 }{2}(2 \Da \textbf{q}^2 +D_r)  -\frac{1 }{2} \sqrt{D_r^2 - 2\textbf{q}^2  (v_0 - \bar{\rho} \zeta) (v_0 -2 \bar{\rho} \zeta)}
\end{split}
\label{computeigenval_eq:4}
\end{equation}

We subsequently expand the eigenvalues up to 2nd order, 
\begin{equation}
\begin{split}
\lambda_{1}&=-( \Da \textbf{q}^2 +D_r)\\
\lambda_{2}&=-1  +\left[-\Da +  \frac{ \left(v_0 - \bar{\rho} \zeta \right)\left(v_0 - 2\bar{\rho} \zeta\right)}{2 D_r}\right]\textbf{q}^2 + \mathcal{O}(\textbf{q}^3)\\
\lambda_{3}&= 0  +\left[-\Da -  \frac{ \left(v_0 - \bar{\rho} \zeta \right)\left(v_0 - 2\bar{\rho} \zeta\right)}{2 D_r}\right]\textbf{q}^2 + \mathcal{O}(\textbf{q}^3)\\
\end{split}
\label{computeigenval_eq:5}
\end{equation}
Out of the three eigenvalues, the only mode which can become unstable (positive) is $\lambda_{3}$. The limit of stability is given by,
\begin{equation}
\zeta = \frac{3}{4}\frac{v_0}{\bar{\rho}} \pm \frac{1}{4\bar{\rho}}\sqrt{v_0^2-16 \Da D_r}
\label{computeigenval_eq:6}
\end{equation}

\section{Computation of parameters}\label{compu_param}

We compute the long-time diffusion coefficient, $\Da$, which will allow us to calculate numerically the normalization factor of the self-propulsion speed, $v^* =4\sqrt{\Da D_{r} }$. To this end, we measure the mean-squared angular displacement of a passive system of particles at $\phi=0.4$ , $\Da = \lim_{t\rightarrow \infty } \frac{1}{4 t}\big\langle( \textbf{r} (t) -  \textbf{r}(0))^2 \big\rangle$. The value of the fit corresponds to $\Da=0.46$, \cref{appx_parameters_fig:1}, which is in good agreement with previous results in the literature \cite{Bialke2013}.

	\begin{figure}[h]
	\centering
	\includegraphics[width=\columnwidth]{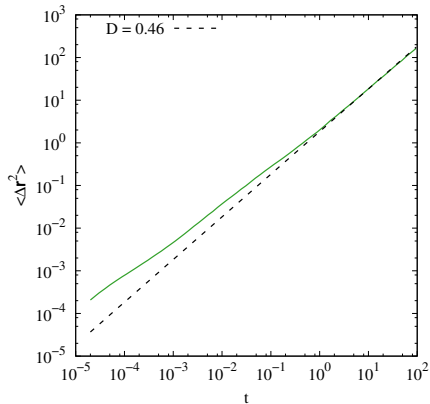}
	\caption{Log-log plot of the mean-squared displacement for a passive system of particles at packing fraction $\phi=0.4$.  }
	\label{appx_parameters_fig:1}
	\end{figure}

\section{Correlation function in the ($r,\theta$)-plane}\label{correl_funct_appx}

 \begin{figure}[h]
	\centering
	\includegraphics[trim=50 50 50 0,width=\columnwidth]{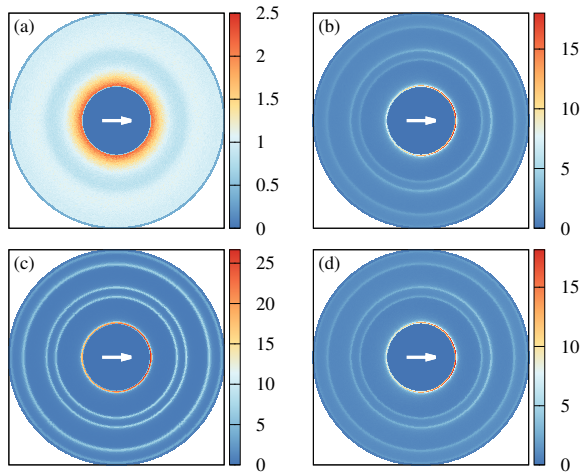}
	\caption{Correlation function $ G(r,\theta)$ of a particle in a system at $\phi=0.4$ for parameters (a) $\mathrm{Pe} = 0$ and $g=0$ (passive suspension); (b) $\mathrm{Pe} = 16.6$ and $g=0$; (c) $\mathrm{Pe} = 16.6$ and ferromagnetic $g/g_f = 0.4$; (d) $\mathrm{Pe} = 16.6$ and nematic $g/g_n = 0.4$. The front of the particle is marked by the white arrow, corresponding to $\theta = 0$.}
	\label{compar_meanfield_microsc_gr_appx_fig:1}
	\end{figure}
	
In the absence of both self-propulsion and local alignment, the correlation function $G(r,\theta)$ is spatially isotropic, as depicted in \cref{compar_meanfield_microsc_gr_appx_fig:1}. It is thus equally probable to find a particle in front than behind the tagged particle, represented by a white arrow. Introducing a finite self-propulsion speed breaks the spatial isotropy, making it more probable to find particles in front than behind the tagged particle, \cref{compar_meanfield_microsc_gr_appx_fig:1} (b). Ferromagnetic alignment further enhances the spatial structure in the plane and the anisotropy between the front and the back of particles, \cref{compar_meanfield_microsc_gr_appx_fig:1} (c). On the contrary, nematic alignment leaves the spatial structure unchanged, \cref{compar_meanfield_microsc_gr_appx_fig:1} (d) with respect to the case with no alignment.

\bibliography{biblio}

\end{document}